\documentclass[twocolumn,superscriptaddress,aps,prl,showpacs,preprintnumbers,amsmath,amssymb,floatfix]{revtex4}

\usepackage[latin9]{inputenc}
\usepackage{color}
\usepackage{amstext}
\usepackage{graphicx}
\usepackage{esint}
\usepackage[unicode=true,
 bookmarks=false,
 breaklinks=false,pdfborder={0 0 1},backref=false,colorlinks=true]
 {hyperref}
 
 \makeatletter
\@ifundefined{textcolor}{}
{%
 \definecolor{BLACK}{gray}{0}
 \definecolor{WHITE}{gray}{1}
 \definecolor{RED}{rgb}{1,0,0}
 \definecolor{GREEN}{rgb}{0,1,0}
 \definecolor{BLUE}{rgb}{0,0,1}
 \definecolor{CYAN}{cmyk}{1,0,0,0}
 \definecolor{MAGENTA}{cmyk}{0,1,0,0}
 \definecolor{YELLOW}{cmyk}{0,0,1,0}
}

\usepackage{dcolumn}
\usepackage{bm}
\usepackage{color}
\usepackage{braket}

\usepackage{mathtools} 

\begin{document}

\title{Realizing Exactly Solvable SU(N) Magnets with Thermal Atoms}

\author{Michael E. Beverland} \affiliation{Institute for Quantum Information \& Matter, California Institute of Technology,  Pasadena, CA 91125, USA}
\author{Gorjan Alagic} \affiliation{Department of Mathematical Sciences, University of Copenhagen}
\author{Michael J. Martin} \affiliation{Institute for Quantum Information \& Matter, California Institute of Technology,  Pasadena, CA 91125, USA}
\author{Andrew P. Koller} \affiliation{JILA, NIST, and Department of Physics, University of Colorado Boulder, CO 80309}
\author{Ana M. Rey} \affiliation{JILA, NIST, and Department of Physics, University of Colorado Boulder, CO 80309}
\author{Alexey V. Gorshkov} \affiliation{Joint Quantum Institute and Joint Center for Quantum Information and Computer Science, NIST/University of Maryland, College Park, MD 20742}

\vskip 0.25cm

\date{\today}

\begin{abstract}
We show that $n$ thermal fermionic alkaline-earth atoms in a flat-bottom trap allow one to robustly implement a spin model displaying two symmetries: the $S_n$ symmetry that permutes atoms occupying different vibrational levels of the trap and the SU($N$) symmetry associated with $N$ nuclear spin states. The symmetries makes the model exactly solvable,
which, in turn, enables the analytic study of dynamical processes such as spin diffusion in this SU($N$) system. We also show how to use this system to generate entangled states that allow for Heisenberg-limited metrology. This highly symmetric spin model should be experimentally realizable even when the vibrational levels are occupied according to a high-temperature thermal or an arbitrary non-thermal distribution.  
\end{abstract}


\maketitle

\pagenumbering{arabic}
The study of quantum
spin models with ultracold atoms \cite{bloch08,bloch12} promises to give crucial insights into a range of equilibrium and non-equilibrium many-body phenomena from 
quantum  spin liquids \cite{balents10}  and  many-body localization \cite{basko06b} to quantum quenches \cite{polkovnikov11,richerme14,jurcevic14} and quantum annealing \cite{das08}. While other approaches exist \cite{wu08b,simon11,pielawa11,schauss12}, the most common approach to implement a quantum 
spin model with ultracold atoms relies on preparing a Mott insulator in an optical lattice, where the internal states of atoms on each site define the effective spin \cite{duan03,bloch08,trotzky08,fukuhara13,greif13,hild14,hart15,brown15}. Virtual hopping processes to neighboring sites and back then give rise to effective superexchange spin-spin interactions. Since the superexchange interactions are typically very weak ($\ll \textrm{kHz}$) \cite{bloch08} (unless the traps are operated near surfaces, which can reduce spacings and increase energy scales \cite{gullans12, romero13, gonzalez2015}), it is a significant challenge in experimental cold atom physics to achieve temperatures and decoherence rates low enough to
access superexchange-based quantum magnetism.

Since ultracold atoms can be prepared in specific internal (i.e. spin) states with extremely high precision, spin temperatures that can be realized are much lower than the experimentally achievable motional temperatures. It is therefore tempting to  circumvent the problem of high motional temperature by constructing a spin model in such a way that the motional and spin degrees of freedom are effectively decoupled. 
We provide a recipe for such a decoupling and hence for realizing spin models with thermal atoms. 

\begin{figure}[b]
\includegraphics[width=0.45\textwidth]{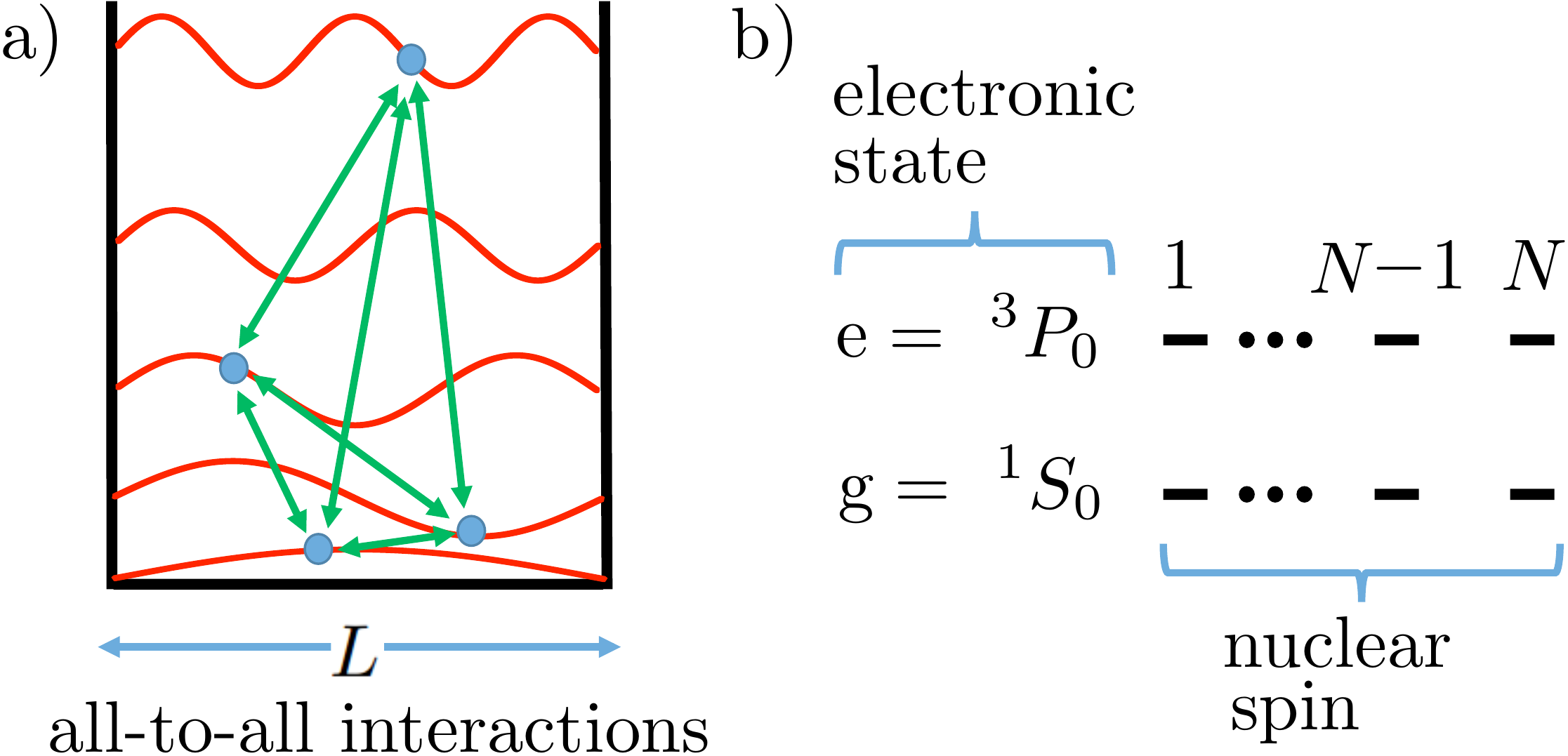}
\caption{(a) Contact interactions between atoms in the orbitals of a one-dimensional infinite square well of width $L$ are all-to-all with equal strength. (b) With nuclear spin $I$, each of the electronic clock states $g$ and $e$ of fermionic alkaline-earth atoms can offer $N$ degenerate
states, with $N \leq 2 I + 1$.}
\label{fig:squarewellwavefunctions}
\end{figure}

The first crucial ingredient for implementing such a spin model is
to depart from second-order superexchange interactions and use contact interactions to first order \cite{gibble09,rey09,yu10,pechkis13,deutsch10,maineult12,hazlett13,martin13,swallows11,koller14}. 
As shown in Fig.~\ref{fig:squarewellwavefunctions}(a), this can be achieved if all atoms sit in different 
orbitals of the same anharmonic trap and 
 remain in these orbitals throughout the evolution, which is a good approximation for weak interactions \cite{martin13,swallows11,gibble09,rey09,yu10}. In that case, the occupied orbitals play the role of the sites of the spin Hamiltonian. However, because of high motional temperature in such systems, every run of the experiment typically yields a different set of populated orbitals and hence a different spin Hamiltonian \cite{martin13}. Thus, unless the dynamics are constrained to states symmetric under arbitrary exchanges of spins \cite{martin13}, every run of the experiment would lead to different spin dynamics. 

The second crucial ingredient to decouple spin and motion is therefore to use an infinite one-dimensional square-well potential as the anharmonic trap, with the motion frozen along the other two directions. 
The interaction terms in the spin Hamiltonian H are proportional to the squared overlap of pairs of distinct sinusoidal orbitals, and are thus all of equal strength.
Therefore $\hat{H}$ is independent of which
orbitals are occupied, leading to spin-motion decoupling and  temperature independent predictions, as well as opening up the possibility of precise control. Moreover,  since $\hat H$  is invariant under any relabeling of the $n$ occupied orbitals, $\hat{H}$ has $S_n$ permutation symmetry. 

Alkaline-earth atoms 
enrich the symmetry. In such atoms, the vanishing electronic angular momentum $J$ in the 
electronic clock states $g = {}^1S_0$ and $e = {}^3P_0$  results in the decoupling of the nuclear spin $I$ from $J$ 
[Fig.~\ref{fig:squarewellwavefunctions}(b)]. This endows $\hat{H}$ with an additional 
$SU(N)$ spin-rotation symmetry, where $N$ can be tuned between $2$ and $2 I + 1$ by choosing the initial state  \cite{gorshkov10, cazalilla09, zhang14, scazza14,pagano14,cappellini14}.
Restricted to $g$, $\hat{H}$ is just the sum of spin-swaps over all 
pairs of occupied orbitals 
and can be 
diagonalized in terms of irreducible representations of the group of symmetries $G=S_n \times SU(N)$. 

 Motional-temperature-insensitive spin models can also be realized using long-range interactions between ions in Paul traps \cite{sorensen99}, Penning traps \cite{richerme14,jurcevic14,britton2012}, and also between molecules \cite{micheli06,barnett06,gorshkov11b,yan13} or Rydberg atoms \cite{schauss12} pinned at different sites of an optical lattice. However, the realization of $SU(N)$-symmetric spin models in such systems
 requires a great deal of fine tuning \cite{gorshkov13}.

Motivated by the exploration of how quantum systems evolve after quantum quenches and whether (or how) they equilibrate and/or thermalize \cite{eisert15}, especially in the presence of long-range interactions \cite{richerme14,jurcevic14}, we first study spin diffusion \cite{sommer11,koschorreck13,yan13} 
in a 
system of $g$ atoms only. 
Due to crucial use of representation-theoretic techniques, our calculations are not only exponentially faster than naive exact diagonalization but also, for $N=2$, yield a closed-form expression for all $n$.
We then present a protocol that employs 
both $g$ and $e$ states 
to create Greenberger-Horne-Zeilinger (GHZ) states \cite{greenberger89}, which could be used to approach the Heisenberg limit for metrology and clock precision \cite{bollinger96}.

\textit{Spin Hamiltonian.}~A single mass-$M$ fermionic alkaline-earth atom (for now, in its ground electronic state $g$) trapped in a 1D spin-independent potential $V(x)$ has real orbitals $\phi_j(x)$ with energies $E_j$ satisfying $\left[ -(\hbar^2/2M) \partial^2/\partial{x^2} + V(x) \right] \phi_j(x) = E_j \phi_j(x)$. The operator $\hat{c}^{\dagger}_{jp}$ creates an atom from the vacuum in $\phi_j(x)$ with nuclear spin state $p \in 1,2, ..., N$. For $n$ identical atoms in the same potential with 
contact $s$-wave interactions, the Hamiltonian is $\hat{H} = \sum_{jp} E_j \hat{c}^{\dagger}_{jp} \hat{c}_{j p} +\sum_{p < q} \sum_{j k j' k' } U_{j k j' k'} \hat{c}^{\dagger}_{jp}  \hat{c}_{j' p} \hat{c}^{\dagger}_{k q} \hat{c}_{k' q}$, where $U_{j k j' k'} = 4 \pi \hbar \omega_\perp a_{gg} \int_{-\infty}^{\infty} d x \phi_j(x) \phi_k(x) \phi_{j'}(x) \phi_{k'}(x)$, $a_{gg}$ is the 3D-scattering length, and a potential with frequency $\omega_\perp$ freezes out transverse motion.

To obtain the desired Hamiltonian, we specialize to a width-$L$ infinite square well $V(x)$, with well-known eigenstates $\phi_j(x) =\sqrt{2/L} \sin(j\pi x/L)$ for $0 \leq x \leq L$, with energy $E_j = (\pi j /L)^2/2M$. Then $U_{j k j' k'}$ is zero unless (i): $(j \pm k) =\pm (j' \pm k')$; to first order in the interaction, we can also set $U_{j k j' k'} \rightarrow 0$ unless $\sum_{jp} E_j \hat{c}^{\dagger}_{jp} \hat{c}_{j p}$ is conserved, which occurs when (ii): $j^2 + k^2 = j'^2 + k'^2$. Both (i) and (ii) are satisfied if and only if $(j',k') =(j,k)$ or $(k',j') =(j,k)$. As the system conserves orbital occupancies, it can be described by a spin model.  
Assuming orbitals are at most singly occupied ($\hat{n}_j = \sum_{p}\hat{c}^{\dagger} _{j p}\hat{c}_{j p} \leq 1$ for all $j$) \footnote{For temperatures far from degeneracy, the probability of multiple occupancy will be small. Alternatively, absence of multiple occupancy is guaranteed by Pauli exclusion for nuclear-spin polarized states.}, the spin Hamiltonian is: 
\begin{equation}
\label{SpinHamiltonian}
\hat{H} = - U\sum_{j < k} \hat{s}_{jk},
\end{equation}
where $\hat{s}_{jk} \equiv \sum_{p q} \hat{c}^{\dagger}_{jp} \hat{c}_{j q} \hat{c}^{\dagger}_{k q}  \hat{c}_{k p}$ swaps spins $j$ and $k$, and the sum is over occupied orbitals. Crucially, $U \equiv 4 \pi  a_{gg} \hbar \omega_\perp/L$ is independent of $j$ and $k$. We dropped a constant 
$\sum_j E_j + n(n-1)U/2$, which will have no effect on spin dynamics. 
For a fixed set of occupied orbitals, $\hat{H}$ has $N^n$ basis states $|p_1, p_2,..p_n\rangle$ with $p_j \in 1, ..., N$.

\textit{Exact eigenenergies and eigenstates.} For $N=2$, the spin-swap can be written in terms of the Pauli operators: $\hat{s}_{jk} = 1/2 + (\hat{\sigma}^x_j \hat{\sigma}^x_k + \hat{\sigma}^y_j \hat{\sigma}^y_k + \hat{\sigma}^z_j \hat{\sigma}^z_k)/2$, allowing Eq.~(\ref{SpinHamiltonian}) to be written as $\hat{H} = - U \left[ \vec{S}^2 + \frac{n}{4}(n-4) \right]$, where $\vec{S} = \frac{1}{2}\sum_j \vec{\sigma}_j $. The eigenstates of $\hat{H}$ for $N=2$ are the well-known Dicke \cite{dicke54} states 
$|S,S_z,k \rangle$, with energies $E(S) = - U \left[ S(S+1) + \frac{n}{4}(n-4) \right]$. The quantum number $k$ labels distinct states with the same $\vec{S}^2$ and $\hat{S}^z$ eigenvalues. We now describe the general case for arbitrary $N$, but defer derivations and detailed explanation to the Supplemental Material \cite{supp}.

Equation~(\ref{SpinHamiltonian}) has two obvious symmetries: permutations in $S_n$ of the $n$ occupied orbitals, and application of the same unitary in $SU(N)$ to all of the spins giving a group $G = S_n \times SU(N)$ of symmetries. From Schur-Weyl duality \cite{fulton91}, we conclude that for each integer partition $\vec{\lambda} = (\lambda_1, \lambda_2,...,\lambda_N)$ such that $\sum_i \lambda_i =n$ and $\lambda_{i+1} \leq \lambda_{i} $, there is a subspace of constant energy $E(\vec{\lambda})$. The $\vec{\lambda}$-subspaces (called irreducible representations of $G$) are orthogonal and span the full Hilbert space. 

\begin{figure}[b]
\includegraphics[width=0.45\textwidth]{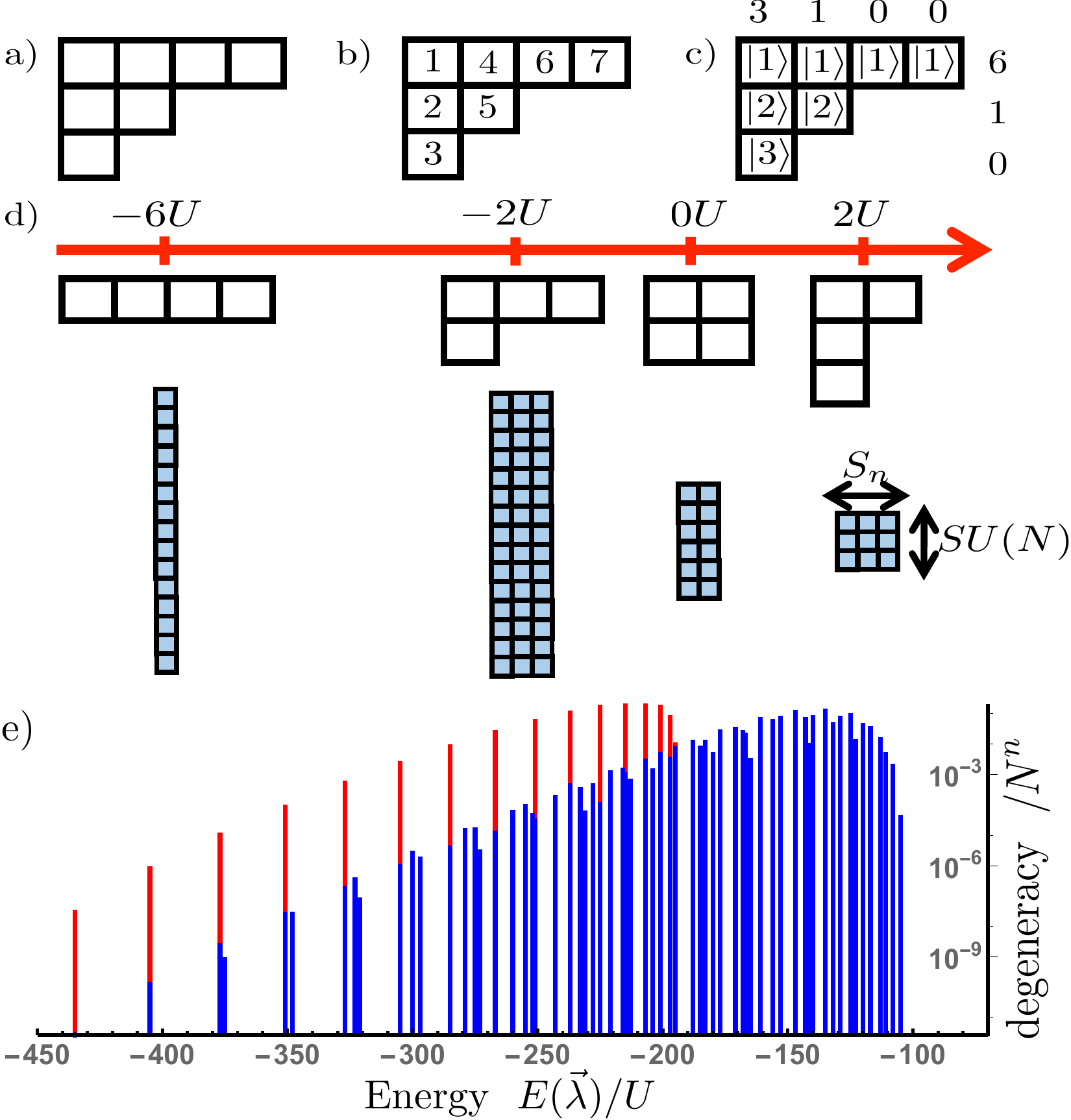}
\caption{(a) A Young diagram $\vec{\lambda} =(4,2,1)$ [with $\vec{\gamma} =(3,2,1,1)$] for $n=7,N=3$. (b) A labeling of boxes in $\vec{\lambda} $ from $1$ to $n$, increasing down columns, starting at the left. (c) Orbitals associated with boxes in row $p$ are put in spin state $|p \rangle$ to form basis state $|T \rangle = |1231211 \rangle$ [spins ordered as in (b)], used to construct eigenstate $|\vec{\lambda} \rangle =| \mathcal{A}\{1 2 3 \} \rangle | \mathcal{A}\{1 2  \} \rangle|1 1 \rangle$ with 
$E(\vec{\lambda})/(-U) = \sum_i {\lambda_i \choose 2}-\sum_j {\gamma_j \choose 2} = 6+1+0 -3-1 - 0 - 0 =3$. (d) The set of all Young diagrams for $n=4$ and $N=3$, with energies above. Below, eigenstates are represented by colored boxes: rotations in $SU(N)$ transform between eigenstates in the same colored column, while permutations in $S_n$ transform between eigenstates in the same colored row. Representative states are found using the prescribed construction to be $\ket{1111}$, $(\ket{12}-\ket{21})\ket{11}$, $(\ket{12}-\ket{21})(\ket{12}-\ket{21})$, and $(\ket{123}+\ket{312}+\ket{231}-\ket{132}-\ket{213}-\ket{321})\ket{1}$, respectively. (e) Spectrum for $n=30$ with $N=2$ (red), and $N=3$ (blue). 
}
\label{fig:YoungDiagramExample}
\end{figure}

A \textit{Young diagram} is a pictorial representation of $\vec{\lambda}$ consisting of a row of $\lambda_1$ boxes above a row of $\lambda_2$ boxes, which is above a row of $\lambda_3$ boxes etc. 
It is also useful to define $\vec{\gamma} = (\gamma_1,\gamma_2,...,\gamma_{\lambda_1})$ as the column heights of the Young diagram $\vec{\lambda}$. Figure~\ref{fig:YoungDiagramExample}(a) shows an example with $n=7$ and $N=3$.

To create an eigenstate in any $\vec{\lambda}$-subspace, first consider the basis state: $|T\rangle \equiv |1 , 2 , ... , \gamma_1 \rangle~ |1 , 2 , ... ,\gamma_2 \rangle ... ~|1 , 2 ,... , \gamma_{\lambda_1} \rangle$, which is chosen by associating orbitals with boxes of the Young diagram as in Fig.~\ref{fig:YoungDiagramExample}(b), and putting those orbitals in spin states as in Fig.~\ref{fig:YoungDiagramExample}(c). We form $|\vec{\lambda}\rangle$ (which is one of many \cite{supp} eigenstates in the $\vec{\lambda}$-subspace) by antisymmetrizing $|T\rangle $ over orbitals associated with boxes in each column of $\vec{\lambda}$:
\begin{equation}
 |\vec{\lambda} \rangle =  | \mathcal{A}\{1 2  ... \gamma_1  \} \rangle | \mathcal{A}\{ 1 2  ... \gamma_2  \} \rangle... |\mathcal{A}\{1 2  ... \gamma_{\lambda_1} \} \rangle,
 \label{Eigenstate}
\end{equation}
where $\mathcal{A}\{...\}$ antisymmetrizes its argument, for example: $| \mathcal{A}\{1 2 3 \} \rangle = |1 2 3\rangle + |31 2\rangle + |2 3 1\rangle -|1 3 2\rangle - |3 2 1\rangle - |2 1 3\rangle$. The normalization constant is fixed by $  \langle \vec{\lambda}  |\vec{\lambda} \rangle = \gamma_1! \, \gamma_2! \,... \gamma_{\lambda_1}!$. We see that the Young diagram associates symmetry with rows and antisymmetry with columns.

From $\hat{H} |\vec{\lambda} \rangle  = E(\vec{\lambda}) |\vec{\lambda} \rangle$ one can prove $E(\vec{\lambda})/(-U) = \sum_{i=1}^N {\lambda_i \choose 2}-\sum_{j=1}^{\lambda_1} {\gamma_j \choose 2}$: the number of ways of choosing two boxes in the same row of $\vec{\lambda}$, minus the number of ways of choosing two boxes in the same column \cite{supp}. This is in line with the intuition that the swap picks up $-U$ for each symmetric pair and $+U$ for each antisymmetric pair in the Young diagram. In terms of $\vec{\lambda}$,
\begin{equation}
\label{EnergyLevelEquation} 
E(\vec{\lambda}) = -\frac{U}{2} \sum_{i=1}^N \left(\lambda_i -2i +1 \right)\lambda_i .
\end{equation}
Figure~\ref{fig:YoungDiagramExample}(d) illustrates the eigenvalues and eigenstates of $\hat{H}$ for the simple case of $n=4$ and $N=3$, along with the corresponding Young diagrams.
There is an equivalence for the $SU(2)$ case between Young diagram $(\lambda_1,\lambda_2)$ and angular momentum quantum number $S$ given by $S = (\lambda_1-\lambda_2)/2 = (2\lambda_1-n)/2$.

\textit{Spin diffusion dynamics.} Spin diffusion is the process by which evolution under a generic spin Hamiltonian causes initially ordered states to diffuse \cite{sommer11,koschorreck13,yan13}. We take initial state $| \psi(0) \rangle =|1 \rangle^{\otimes m_1} |2 \rangle^{\otimes m_2} ...  |N \rangle^{\otimes m_N}  $. 
Note any computational basis state can be changed to this form by reordering occupied orbitals. We consider the time evolution of observable $\hat{Q} = \sum_{j=1}^{m_1} |1 \rangle_j \langle 1 |_j$: the number of the first $m_1$ orbitals in spin-state $|1\rangle$.  This is the simplest observable capturing the broken symmetry of the initial state. The expectation of $\hat{Q}$  evolves according to: $Q(t) \equiv \langle \psi(0)| e^{i\hat{H}t} \hat{Q} e^{-i\hat{H}t} | \psi(0) \rangle$, omitting $\hbar$ where convenient from here on.

\begin{figure}[tbh]
\includegraphics[width=0.45\textwidth]{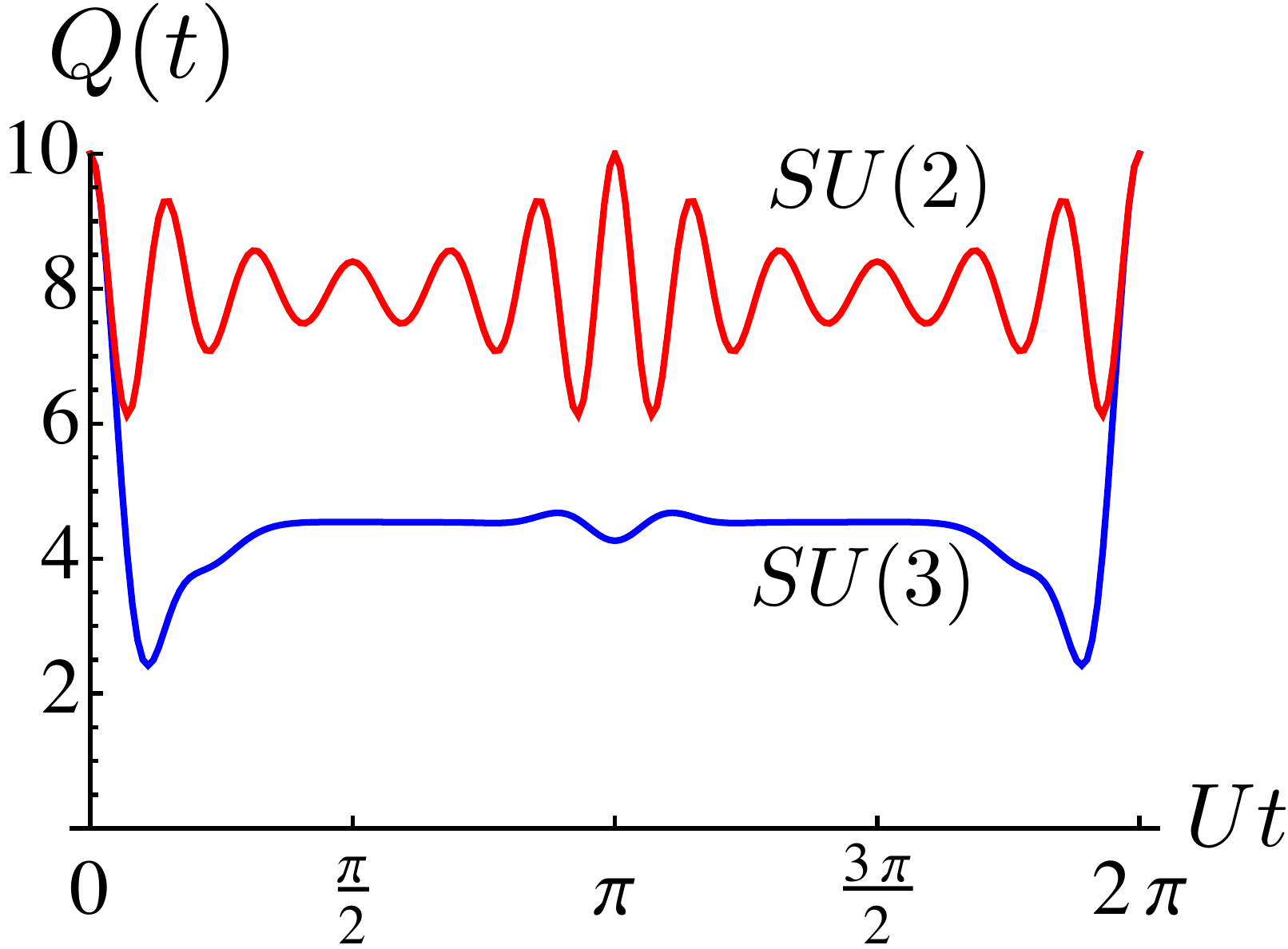}
\caption{Exact time evolution of $\hat{Q} = \sum_{j=1}^{10} |1 \rangle_j \langle 1 |_j$, which counts the number of the first ten orbitals in spin state $|1\rangle$. Two initial states are compared: $|1 \rangle ^{ \otimes 10} |2 \rangle ^{ \otimes 20} $ for $SU(2)$ and $|1 \rangle ^{ \otimes 10} |2 \rangle ^{ \otimes 10} |3 \rangle ^{ \otimes 10} $ for $SU(3)$. 
The initial evolution is similar, but more $|1 \rangle$ states diffuse out of the first ten orbitals for $SU(3)$ later on. Since all $E(\vec{\lambda})$ are integer multiples of $U$, complete revival occurs at $Ut = 2 \pi$. In the $SU(2)$ case, the oscillation is dominated by the smallest $S$ in Eq.~(\ref{Nequals2TimeEvolutionArb}). This is consistent with the fact that for fixed $S_z$, the size of the eigenspaces decreases with $S$, causing overlap to be larger with subspaces of small $S$ generically.}
\label{fig:SpinDiffusion}
\end{figure}

Calculating $Q(t)$ for a generic Hamiltonian requires matrix diagonalization, which scales exponentially with $n$ (for fixed $N$). Using the symmetry of Hamiltonian 
(\ref{SpinHamiltonian}) and the Wigner-Eckart theorem for $SU(N)$, we obtain an explicit sum (see Eq.~(S11) in Ref.~\cite{supp}) for $Q(t)$ in terms of Clebsch-Gordan and recoupling coefficients. For the case of $N=2$, with initial state of $m_1 =m$ spin up and $m_2 =n-m$ spin down orbitals, using well-known closed forms for the Clebsch-Gordan and recoupling coefficients:
\begin{equation}
\label{Nequals2TimeEvolutionArb}
Q(t) = m + \sum_{\mathclap{S=|n-2m|/2+1}}^{n/2} \gamma(S) [\cos{(2SUt)}-1],
\end{equation}
where $\gamma(S) = \frac{4S^2\!-\!(n\!-\!2m)^2}{4S}  \binom{n}{n/2+S}/\binom{n}{n-m} $. For $N>2$, closed forms for the required coefficients are not known to the authors, but can be calculated efficiently using standard algorithms as in Ref.~\cite{alex11}. In Fig.~\ref{fig:SpinDiffusion}, we compare the evolution of the same operator and total particle number for initial states with $N= 2$ spin states and $N=3$ spin states. The oscillations are much less pronounced and spin diffusion 
occurs more fully ($Q$ drops lower)  for the latter state. With this model, looking at times away from the multiples of the revival time $2 \pi/U$, one could study apparent near-equilibration of some observables (such as $Q$ in the $N=3$ case) acting on the first $m_1$ spins. Perturbations could be added to the system to remove revivals and  potentially allow for the thermalization of the first $m_1$ spins. 

\textit{GHZ state preparation.} Highly entangled states could lead to short-term 
applications in metrology \cite{bollinger96,sackett00}, and long-term applications in quantum information \cite{nielsen00,dutta13}. It is particularly timely to design ways for implementing entanglement-assisted -- and hence more accurate -- clocks with alkaline-earth atoms \cite{gil14,olmos13} since such atoms recently gave rise to the world's best clock and have nearly approached the quantum projection noise limit for unentangled atoms \cite{bloom14,nicholson15}. We now show our system offers a natural way to produce metrologically relevant entanglement (in the form of GHZ states) in alkaline-earth clock experiments. It is the experimental realization of quantum spin models in alkaline-earth clock experiments \cite{martin13} and the potential application of these spin models to improve the clocks that motivated this work.
  
\begin{figure}[t]
\includegraphics[width=0.47\textwidth]{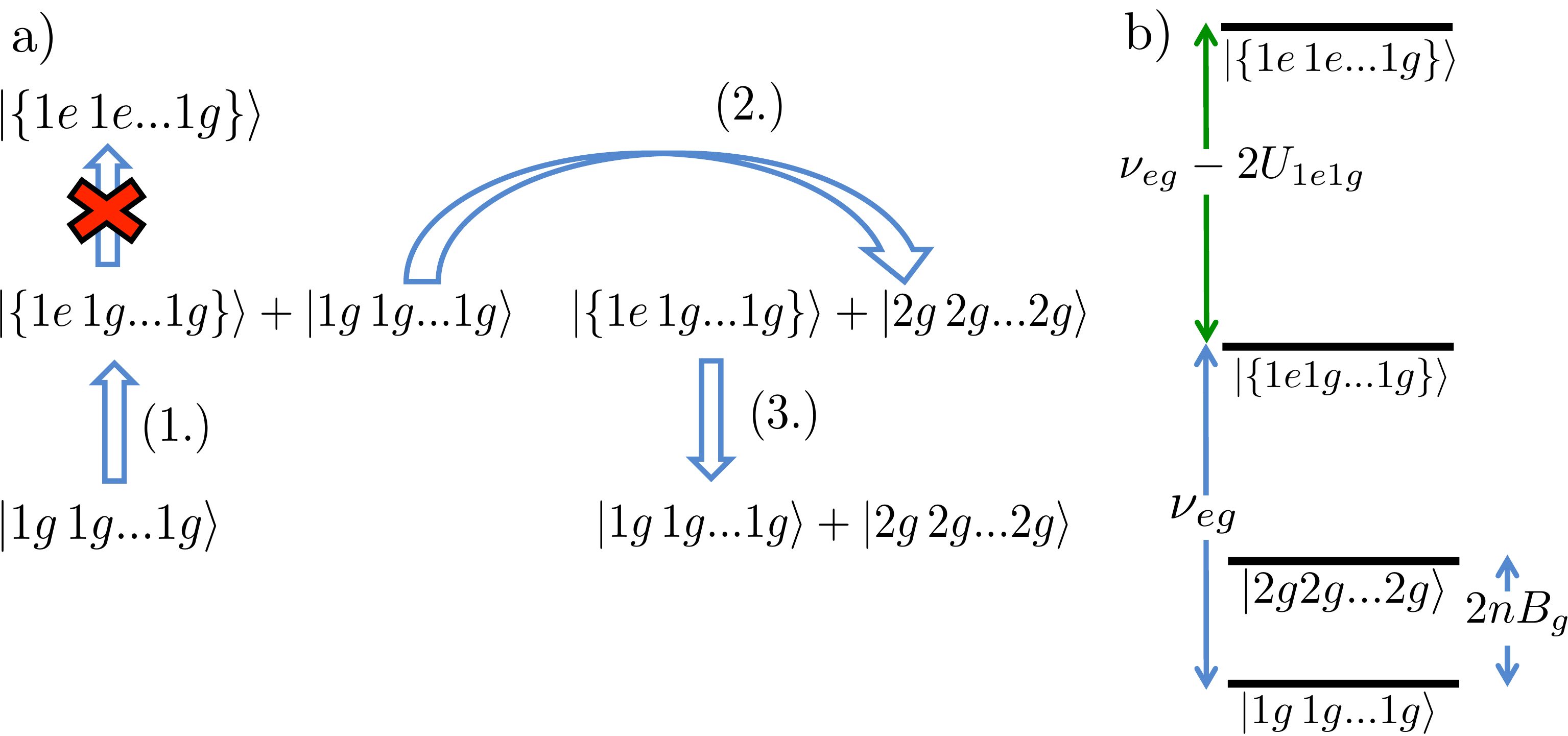}
\caption{(a) System prepared in $| 1g 1g .. 1g \rangle$. Spatially inhomogeneous pulse (1.) results in equal superposition of this state and $|\{ 1e 1g .. 1g \} \rangle$, 
containing one $e$ atom. An interaction blockade prevents coupling to states with two $e$ atoms. Pulse (2.) flips the spins of the all-$g$ state. The initial pulse is reversed in pulse (3.), resulting in the GHZ state. (b) Relevant energy levels of the Hamiltonian with $e$ and $g$ states and the magnetic field. 
Note that pulses (1.) and (3.), which involve states $| 1g 1g .. 1g \rangle$ and $|\{ 1e 1g .. 1g \} \rangle$, do not couple to state $|\{ 1e 1e .. 1g \} \rangle$ since there is a blockade of $2U_{1e1g}$. Similarly, during pulse (2.), blockade prevents excitation of $|\{ 1e 1g .. 1g \} \rangle$.}
\label{fig:GHZProtocol2}
\end{figure}

To create a GHZ state, we allow atoms in the excited electronic state $e$ with energy $\omega_{eg}$ above the ground electronic state $g$ [see Fig.\ \ref{fig:squarewellwavefunctions}(b)]. First assume $N=2$. An applied magnetic field adds Zeeman spin-splittings $B_g \neq B_e$ \cite{boyd06} to both $g$ and $e$ states. 
To first order in the interaction strength, the spin Hamiltonian is \cite{supp}:
\begin{eqnarray}
\label{EandGatoms}
 \hat{H} &=& \hat{H}_{sp} +\sum_{\alpha < \beta}U_{\alpha \beta} \left( \hat{n}_{\alpha} \hat{n}_{\beta} - \sum_{j\ne k} \hat{c}^{\dagger}_{j \alpha} \hat{c}_{j \beta} \hat{c}^{\dagger}_{k \beta} \hat{c}_{k \alpha} \right).~
\end{eqnarray}
The single-particle Hamiltonian is $\hat{H}_{sp} = \omega_{eg} \hat{n}_{e} + B_g(\hat{n}_{1g}-\hat{n}_{2g})+ B_e( \hat{n}_{1e}-\hat{n}_{2e}) $, the sum $\alpha< \beta$ is over distinct pairs of $1g$, $1e$, $2g$ and $2e$. Constants $U_{\alpha \beta}$ are derived in terms of (electronic-state dependent) scattering lengths \cite{supp}. Note that $\hat{n}_{1g}$, $\hat{n}_{2g}$, $\hat{n}_{1e}$ and $\hat{n}_{2e}$ are separately conserved by Hamiltonian 
(\ref{EandGatoms}). As shown in Fig.~\ref{fig:GHZProtocol2}, to create the $n$-particle GHZ state $(|1g 1g.. 1g \rangle + |2g 2g.. 2g \rangle )$ from $|1g 1g.. 1g \rangle$, three consecutive pulses should be applied:
\begin{enumerate}
\item Spatially inhomogeneous, weak, many-body $\pi/2$ pulse $e^{-i\nu_{eg}t}~\sum_{j} \Omega^{eg}_j (|1e \rangle_j \langle 1g|_j + |2e \rangle_j \langle 2g|_j ) + h.c.$ with frequency $\nu_{eg} = \omega_{eg} + (B_e-B_g) + n U_{1e 1g}$. 
\item Spatially uniform, weak, single-atom $\pi$ pulse $e^{-i\nu_{12}t}\Omega^{12}\sum_{j} ( |2g \rangle_j \langle 1g|_j +| 2e \rangle_j \langle 1e|_j  ) + h.c.$ with frequency $\nu_{12} = 2 B_g$.
\item Pulse 1, but for pulse area $\pi$, not $\pi/2$. 
\end{enumerate}
The frequency of the first pulse picks out an effective two-level system consisting of $| 1g 1g .. 1g \rangle$ and $|\{ 1e 1g .. 1g \} \rangle \propto \sum_{jp} (\Omega^{eg}_j-\bar{\Omega}^{eg}) |1e \rangle_j \langle 1g|_j |1g 1g .. 1g \rangle$ (we defined $\bar{\Omega}^{eg} \equiv \sum_{j} \Omega^{eg}_j/n$.). The pulse must be spatially inhomogeneous to make $\Omega^{eg}_j$ $j$-dependent and to be able to access eigenstates with interaction-dependent energies (i.e.\ not fully symmetric eigenstates). The precise form of the inhomogeneity is unimportant, as all $n-1$ non-symmetric states with a single $e$ atom are degenerate in $\hat{H}$ due to its $S_n$ symmetry. We use curly brackets to signify linear combinations of $| 1e 1g .. 1g \rangle$ and permutations. No state $|\{ 1e 1e .. 1g \} \rangle$ is coupled by pulse 1 because the first $e$ atom blockades the addition of another by energy $2U_{1e1g}$ \cite{supp}. The second pulse has no effect on $|\{ 1e 1g .. 1g \} \rangle$ because the $e$ atom blockades transition to any state $|\{ 1e 2g .. 1g \} \rangle$. The final pulse does not affect the $| 2g 2g .. 2g \rangle$ state because the pulse is off-resonant by energy of order $(B_e-B_g)$ \cite{supp}. Note that although the precise form of the inhomogeneity in the first pulse is unimportant, the final pulse and the first pulse must have the same inhomogeneity. Since all three pulses rely on blockade, each pulse must take time $\gg 1/U$. 
Curiously, the fact that the interactions in our spin model have effectively infinite range makes our spins analogous to long-range interacting Rydberg atoms, for which a similar protocol exists for generating maximally entangled states \cite{saffman09}. We have designed the protocol to have at most one $e$ atom 
at any time, which avoids the potential problem of inelastic $e$-$e$ collisions \cite{traverso09}, while $g$-$e$ losses are negligible \cite{bishof11b, zhang14}.  

For integer $m$ such that $N \geq 2^m$, $m$ GHZ states can be created provided one has sufficient control \cite{gorshkov09} over the nuclear spin states coupled by the pulses \cite{supp}. 
Several GHZ states can be used to create a single GHZ state of better fidelity via entanglement pumping \cite{aschauer05,gorshkov09}.

\textit{Experimental Considerations.}  We use the example of ${}^{87}$Sr to describe how to experimentally access the physics we discuss in this work. 

The key requirements of this proposal are as follows. Firstly, the $x$ and $y$ degrees of freedom must be frozen, forming a 1D interacting system along the $z$ direction. Secondly, $U=(4 \pi a_{gg} \hbar \omega_\perp)/L$ should be less than the single-particle energy separations, the smallest of which is $3 \hbar^2(\pi /L)^2/M$, ensuring the validity of the first-order perturbation theory in our derivation of Eq.~(\ref{SpinHamiltonian}). This constrains the relative sizes of $L$ and $\omega_{\perp}$. Thirdly, variations in $U_{jkjk}$, with standard deviation $\Delta U$, give rise to variations in eigenergies $ \sim n \Delta U$ (see Supplemental Material \cite{supp}). Therefore, we also require $\Delta U/U < 1/n$.

\begin{figure}[tbh]
\includegraphics[width=0.48\textwidth]{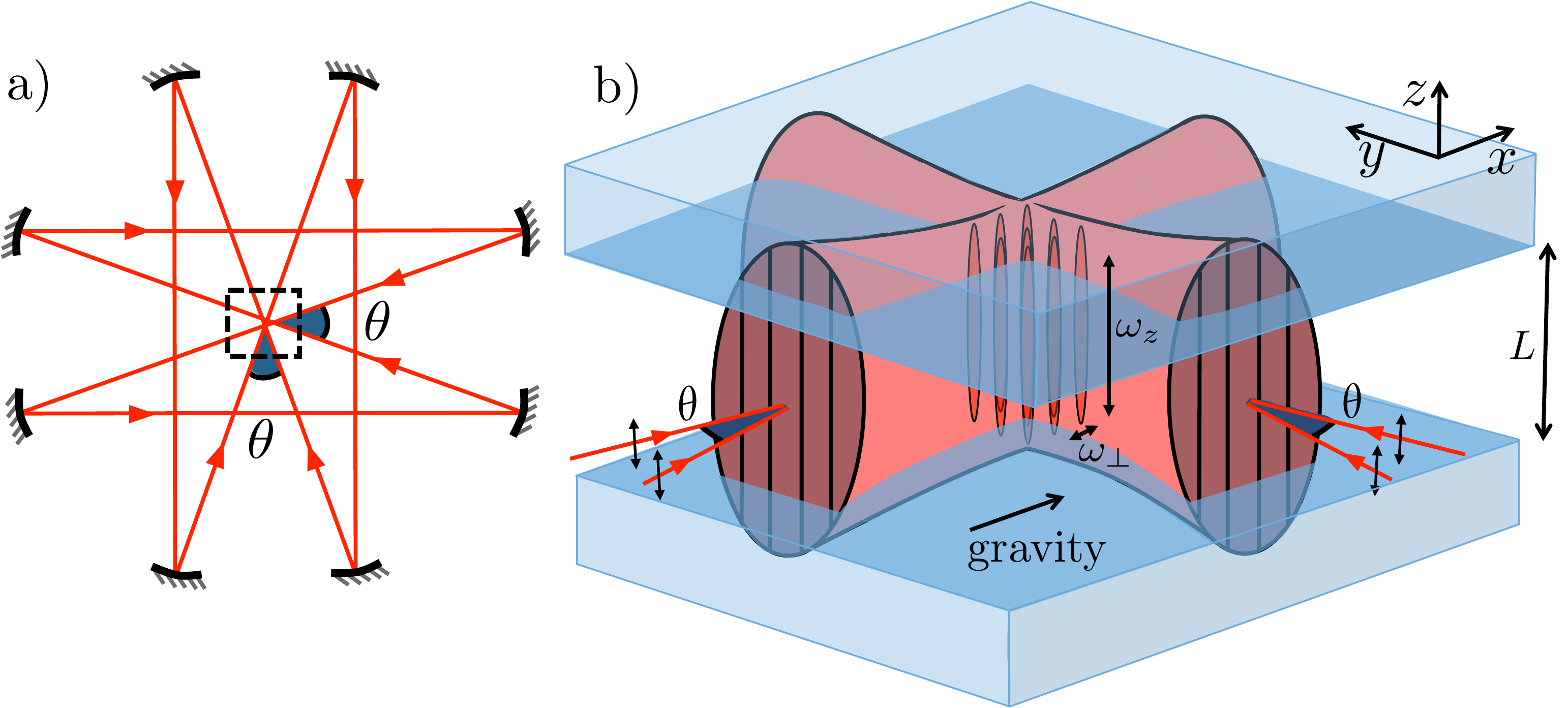}
\caption{Layout of suggested experimental implementation. a) A bow tie beam arrangement of two pairs of beams aimed at a vacuum chamber. In each pair, the two beams have different $k$ vector directions of $\theta = 30^\text{o}$, forming an in-plane standing wave perpendicular to that pair's net $k$ vector direction. The pair of perpendicular standing waves forms an attractive lattice. b) The two-dimensional lattice of attractive-potential tubes forms with transverse vibrational frequency $\omega_{\perp}$ and lattice constant $\Delta x$. The finite beam width results in a weak potential in the $z$ direction with vibrational frequency $\omega_z$. Gravity is in the beam plane to avoid a potential gradient along the tubes. Blue-detuned light outside the central region of width $L$ forms caps for the tubes. Following the Supplemental Material \cite{supp}, we obtain $\omega_{\perp} \simeq 2 \pi \times 10$~kHz, $\Delta x \simeq 3$~$\mu$m, $\omega_z \simeq 2 \pi \times  100$~Hz, and $L \simeq 10$~$\mu$m.
} 
\label{fig:ExperimentalSetup}
\end{figure}

To meet these requirements, we 
propose an optical lattice potential formed by two magic-wavelength (813 nm) \cite{ye08} orthogonal standing waves in $x$ and $y$. This could be achieved with a pair of angled beams \cite{nelson07} for each standing wave, in bow tie configuration [see Fig.~\ref{fig:ExperimentalSetup}]. 

An additional blue-detuned optical potential at 394~nm, the Sr blue magic wavelength, is applied to form approximate 1D square wells from the resulting tubes. The potential could be formed from a projected image of a Gaussian beam with waist 30~$\mu$m and total power 400~mW screened in the center by a rectangular mask of width $L$ = 10~$\mu$m. Imperfect cap potentials, along with a finite curvature of the flat potential, contribute to $\Delta U$ and are analyzed in the Supplemental Material \cite{supp}.

With these parameters, and $a_{gg} = 5.1$ nm \cite{Escobar08},  one obtains $U/\hbar =(4 \pi a_{gg} \omega_\perp)/L \approx 2 \pi \times 10$ Hz, and should be able to meet all three of the aforementioned key requirements with  $\lesssim 20$ atoms in a single tube. Further details are included in the Supplemental Material \cite{supp}. Such values of $U_{\alpha \beta} \sim U$ \cite{zhang14} can potentially allow for the preparation of the GHZ state on a time scale comparable to the $\sim\!1$s experimental cycle time for state-of-the-art clocks \cite{bloom14}, and may thus provide a practical advantage over the use of unentangled atoms. 

To observe spin diffusion, the initial state could be formed by cooling a spin-polarized system to the limit where the lowest $n$ orbitals are occupied. One could potentially consider taking advantage of large $N$ for better cooling \cite{hazzard12,taie12}. One coud address different orbitals either spatially with spin-changing pulses which only couple to certain orbitals (for example using pulses focused on the center of the well and hence decoupled from orbitals that vanish there), or energetically by temporarily transferring atoms to another electronic state subject to a different potential. To observe spin diffusion with thermal atoms, one could rely on the fact that about half of the occupied orbitals are odd, and the other half are even, which becomes statistically more accurate for larger $n$. It is possible to address only the even orbitals by using a beam focused at the center of the well, since the odd orbitals vanish there. This could be extended to larger $N$ by using additional beams focused on other points in the well. 

\textit{Outlook.} The proposed system opens a wide range of research and application avenues beyond those discussed above. 
For the case of $N=2$, our $S_n \times SU(N)$-symmetric Hamiltonian can be used for decoherence-resistant entanglement generation \cite{rey08}, a method whose generalization to $N > 2$ we postpone to future work. 
Furthermore, by comparing with the exact solutions presented here or those derived in the limit of strong interactions \cite{volosniev15,deuretzbacher14} one could verify the performance of the proposed experimental system as a quantum simulator. The system can then be used to reliably study more general regimes where complexity theory might rule out efficient classical solutions. 
In particular, 
deviations from the square-well potential will break $S_n$ [but not $SU(N)$] symmetry. This will for example lift the degeneracy of the most antisymmetric spin state (highest energy eigenspace for $U>0$). Depending on how this degeneracy is lifted, exotic many-body states 
might arise \cite{cazalilla14, rey14}. 

Finally, thanks to its high $S_n \times SU(N)$ symmetry, the present system allows one to implement powerful quantum information protocols, such as the density matrix spectrum estimation protocol of Keyl and Werner \cite{keyl01,beverland15}.

\begin{acknowledgments}
We thank S.\ Jordan, J.\ Haah, J.\ Preskill, K.\ Hazzard,  G.\ Campbell, E.\ Tiesinga, and D.\ Barker for discussions. 
This work was supported by NSF IQIM-PFC-1125565, NSF JQI-PFC-0822671, NSF JQI-PFC-1430094, NSF JILA-PFC-1125844, NSF-PIF, NIST, ARO, ARL, ARO-DARPA-OLE, AFOSR, AFOSR MURI, and the Lee A. DuBridge and Gordon and Betty Moore foundations. APK was supported by the Department of Defense through the NDSEG program. MEB and AVG acknowledge the Centro de Ciencias de Benasque Pedro Pascual for hospitality. 
\end{acknowledgments}

\begin{widetext}

\renewcommand{\thesection}{S\arabic{section}} 
\renewcommand{\theequation}{S\arabic{equation}}
\renewcommand{\thefigure}{S\arabic{figure}}
\setcounter{equation}{0}
\setcounter{figure}{0}

\section{Eigenstates and energies of the Hamiltonian}

In this Section, we present the details behind the derivation of the eigenstates and the energies of the Hamiltonian given in Eq.\ (1) of the main text. In particular, we compute the degeneracy of the ground state for $U > 0$ and $U < 0$. As in the main text, we use $n$ and $N$ to mean the number of atoms, and number of nuclear spin states per atom respectively.

Define $\hat{U}(\hat{V},\sigma)$ which permutes occupied orbitals by $\sigma \in S_n$ and implements the spin rotation $\hat{V} \in SU(N)$:
\begin{equation} \label{unitarydefn}
\hat{U}(\hat{V},\sigma) ~ |p_1 \rangle |p_2 \rangle  ... |p_n \rangle ~~~ \equiv ~~~ \hat{V}|p_{\sigma^{-1}(1)} \rangle \hat{V}|p_{\sigma^{-1}(2)} \rangle ... \hat{V}|p_{\sigma^{-1}(n)} \rangle.
\end{equation}
These unitaries (for all $\hat{V} \in SU(N)$ and $\sigma \in S_n$) form a well-understood representation of the group $G = S_n \times SU(N)$. Each such unitary commutes with $\hat{H}= - U \sum_{j \neq k} \hat{s}_{jk}$, where for clarity we dropped all constants from Eq.~(1). Irreps of $SU(N)$ and $S_n$ are uniquely labeled by Young diagrams $\vec{\mu}$ and $\vec{\nu}$, respectively, which satisfy different conditions: $\vec{\mu} = (\mu_1, \mu_2,..\mu_N)$, whereas $\sum_i \nu_i =n$. Each irrep of the product group $G = S_n \times SU(N)$ is the tensor product of an irrep of $SU(N)$ and an irrep of $S_n$ and is therefore uniquely labeled by a pair $(\vec{\mu},\vec{\nu})$. A consequence of Schur-Weyl duality is that representation
(\ref{unitarydefn}) block-diagonalizes into exactly one copy of each irrep of $G$ satisfying $\vec{\mu} = \vec{\nu}$, and no other irreps \cite{bacon07,fulton91}. Therefore for each Young diagram $\vec{\lambda} = (\lambda_1, \lambda_2,..,\lambda_N)$ such that $\sum_i \lambda_i =n$, there is a subspace of constant energy $E(\vec{\lambda})$. One can form an unnormalized projection operator $\hat{\Pi}_{L(\vec{\lambda})}$ into the $\vec{\lambda}$ subspace \cite{fulton91}:
\begin{equation}
\hat{\Pi}_{L(\vec{\lambda})} = \sum_{\substack{\text{$c \in \text{col}(T)$}\\\text{$r\in \text{row}(T)$}}} \text{sgn}(c) ~~ \hat{U}(\hat{I},c)~ \hat{U}(\hat{I},r).
\end{equation}
Here, $L(\vec{\lambda})$ is the labeling of boxes in the Young diagram $\vec{\lambda}$ from $1$ to $n$ as shown in Fig.~2(b) in the main text, and $\text{row}(L)$ ($\text{col}(L)$) is the group of all permutations of the numbers $1$ to $n$ that preserve the contents of rows (columns) of $L(\vec{\lambda})$. Applying $\hat{\Pi}_{L(\vec{\lambda})}$ to any state that it does not annihilate returns an eigenstate of energy $E(\vec{\lambda})$. For concreteness we use $|T\rangle \equiv |1 , 2 , ... , \gamma_1 \rangle~ |1 , 2 , ... ,\gamma_2 \rangle ... ~|1 , 2 ,... , \gamma_{\lambda_1} \rangle$, where we also define $\vec{\gamma} = (\gamma_1,\gamma_2,...,\gamma_{\lambda_1})$ as the column heights of the Young diagram $\vec{\lambda}$. For each $\vec{\lambda}$ we obtain an explicit eigenstate: $|\vec{\lambda} \rangle = \hat{\Pi}_{L(\vec{\lambda})} |T\rangle$ as in Eq.~(2) of the main text. Now we describe how to obtain the eigenvalue $E(\vec{\lambda})$ such that:
\begin{equation}
\hat{H} | \vec{\lambda} \rangle = E(\vec{\lambda}) | \vec{\lambda} \rangle.
\end{equation}
Premultiplying by $\langle T|$ we obtain: $E(\vec{\lambda}) = \langle T |\hat{H} | \vec{\lambda} \rangle = - U \sum_{j \neq k} \langle T |\hat{s}_{jk} | \vec{\lambda} \rangle$, noting that $\langle T | \vec{\lambda} \rangle =1$. For $j,k$ in the same column of the labeled Young diagram $L(\vec{\lambda})$, we know that $\hat{s}_{jk} | \vec{\lambda} \rangle = -  | \vec{\lambda} \rangle$. Similarly for $j,k$ in the same row of $L(\vec{\lambda})$ we have  $\langle T | \hat{s}_{jk} = \langle T |$. Thus pairs $(j,k)$ in columns contribute $-1$ to $E(\vec{\lambda})$ and pairs $(j,k)$ in rows contribute $+1$. The number of such pairs can be counted, hence:
 \begin{equation}
E( \vec{\lambda})/(-U)  =\sum_{i=1}^N {\lambda_i \choose 2}-\sum_{j=1}^{\lambda_1} {\gamma_j \choose 2} + \sum_{\{j \neq k\}_{\text{diagonal}}} \langle T |\hat{s}_{jk} | \vec{\lambda} \rangle,
\end{equation}
The swap $\hat{s}_{jk}$, where $j$ and $k$ are neither in same column nor in same row in $L(\vec{\lambda})$, can always be written as $\hat{s}_{jk} = \hat{s}_{jm} \hat{s}_{km} \hat{s}_{jm} = \hat{s}_{km}\hat{s}_{jm}\hat{s}_{km}$, where $m$ is chosen such that $(j, m)$ and $(k, m)$ lie in a row and a column of $L(\vec{\lambda})$, respectively (it suffices to consider the case $j>k$). Therefore, $\langle T |\hat{s}_{jk} | \vec{\lambda} \rangle = \langle T | \hat{s}_{km} \hat{s}_{jm} | \vec{\lambda} \rangle = - \langle T |\hat{s}_{km}\hat{s}_{jm} | \vec{\lambda} \rangle  = 0$, implying $E(\vec{\lambda})/(-U) = \sum_{i=1}^N {\lambda_i \choose 2}-\sum_{j=1}^{\lambda_1} {\gamma_j \choose 2}$.

The dimensions of each block can be calculated using the standard hook-length formulae \cite{sagan00} for any given Young diagram $\vec{\lambda}$. In particular, the ground-state spaces for $U>0$ (ferromagnetic interaction) and $U<0$ (antiferromagnetic interaction) are $\vec{\lambda}_{F} = (n, 0, 0, ..., 0)$ and $\vec{\lambda}_{AF} = (n/N, n/N, ..., n/N)$ and have dimensions $ D_{F} $ and $ D_{AF} $, respectively:
\begin{eqnarray}
 D_{F}  =  \frac{(n+N-1)!}{n!~(N-1)!} ,~~~~~~~~~~~~ D_{AF}  =  \frac{n!}{[(n/N)!]^N} \prod_{i=1}^{N-1}  \frac{i!}{[n/N+i]} .
\end{eqnarray}

\section{Derivation of spin-diffusion dynamics} 

In this Section, we present the derivation of the spin-diffusion dynamics, first for $N=2$ [i.e.\ Eq.\ (4) in the main text] and then for general $N$.

We are concerned with observable $\hat{Q} = \sum_{j=1}^{m_1} |1 \rangle_j \langle 1 |_j$. In this section, we use the notation that for any operator $\hat{A}$,  $A(t) \equiv \langle \psi(0) | e^{i \hat{H} t} \hat{A}  e^{-i \hat{H} t} | \psi(0) \rangle$, where $| \psi(0) \rangle =|1 \rangle^{\otimes m_1} |2 \rangle^{\otimes m_2} ...  |N \rangle^{\otimes m_N}  $. As most readers are assumed to be familiar with spin-$1/2$ systems, we outline the $N=2$ case first before covering the general case more abstractly. 

For $N=2$, we can choose the angular momentum (Dicke) basis to span the Hilbert space: $|S, S_z,k \rangle$, which diagonalizes the Hamiltonian: $\hat{H} |S, S_z,k \rangle = - U S(S+1)|S, S_z,k \rangle$ (dropping a constant energy). The initial state is $|\psi(0) \rangle = \ket{ \uparrow }^{\otimes m} \ket{ \downarrow }^{\otimes n-m}$ where we used $\ket{\uparrow }$ and $\ket{\downarrow }$ in place of $|1 \rangle$ and $|2 \rangle$. This state can be understood as a tensor product of two Dicke states on subsets of spins: $|\psi(0) \rangle = | m/2,m/2 \rangle \otimes | (n-m)/2, -(n-m)/2  \rangle$, where there is no need for a $k$ quantum number 
since states with $|S_z| = S$ 
have no additional degeneracy. The tensor product of two angular momentum states can be written as a sum of ``total'' angular momentum states: $|\psi(0) \rangle = \sum_{S} C(S) |S,S_z \!\!=\!\! m \!-\! n/2, \alpha(S) \rangle$, where $C(S)$ is a Clebsch-Gordan coefficient, and $\alpha(S)$ represents the fact that $ |S,S_z \!\!=\!\! m \!-\! n/2, \alpha(S) \rangle$ is some specific linear combination of Dicke states with the same $S$ and $S_z$, but different $k$'s. Hence, $Q(t) = \sum_{S,S'} C(S')^* C(S) e^{iUt\left[S(S+1) - S'(S'+1) \right]} \langle S',S_z,\alpha(S')| \hat{Q} | S,S_z,\alpha(S) \rangle$.  Note that $\hat{Q}=m\hat{I} + \hat{S}^z_{m}$ with $\vec{S}_{m} = \sum_{j=1}^{m}\vec{S}_j$, and $\hat{S}^z_{m}$ is the $0$-component of the $(S=1)$-spherical tensor $\hat{\mathbb{T}} \equiv \{ \hat{S}^{-1}_{m}, \hat{S}^z_{m}, \hat{S}^{+1}_{m} \}$, with $\hat{S}^{\pm1}_{m} = \mp (\hat{S}^x_{m} \pm i \hat{S}^y_{m})/\sqrt{2}$. We first apply the Wigner-Eckart theorem to write the matrix element in terms of the reduced matrix element and a Clebsch-Gordan coefficient. Then, since $\hat{\mathbb{T}} \equiv \hat{\mathbb{T}}_{m}\otimes \hat{I} $ acts only on the first $m$ spins, we rewrite \cite{Rose57,brown2003} the reduced matrix element on the full system in terms of one on the first $m$ spins and a recoupling coefficient:
\begin{eqnarray}
\label{UsefulIdentitySU2}
\!\!\! \!\!\!  \langle S',S_z', \alpha(S') | \hat{Q} | S,S_z, \alpha(S) \rangle &=& m \delta_{S,S' } + \langle m/2 || \hat{\mathbb{T}}_L|| m/2 \rangle~ \left\{ \begin{array}{ccc}
 1 & m/2 & m/2\\
 (n-m)/2 & S' & S \end{array} \right\} \left( \langle 1,0| \otimes \langle S,S_z| \right)~|S',S_z' \rangle,
\end{eqnarray} 
where $\left( \langle 1,0| \otimes \langle S,S_z| \right)~|S',S_z' \rangle$ is a Clebsch-Gordan coefficient and $\langle m/2 || \hat{\mathbb{T}}_L|| m/2 \rangle$ is the reduced matrix element of $\hat{\mathbb{T}}_L$ on the $S=m/2$ state of the first $m$ spins. The recoupling coefficient  $
\left\{ \begin{array}{ccc}
 S_A & S_B & S_{AB}\\
 S_C & S & S_{BC} \end{array} \right\} \equiv \langle S, S_z, (S_{AB}, S_C) | S, S_z, (S_A,S_{BC}) \rangle
$ is the overlap between two states of given $S$ and $S_z$ formed from the tensor product of three subsystems with $S_A$, $S_B$ and $S_C$ in two different ways: by combining $A$ and $B$ to form $S_{AB}$ first, and by combining $B$ and $C$ to form $S_{BC}$ first. Substitution of the Clebsch-Gordan and recoupling coefficients into the matrix element gives Eq.~(4) in the main text.

\begin{figure}[t]
\includegraphics[width=0.95\textwidth]{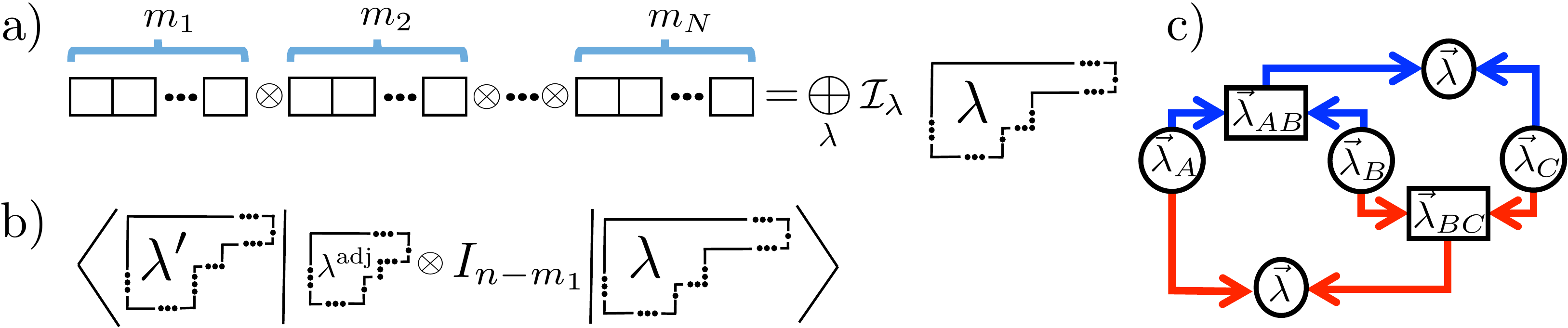}
\caption{(a) Initial state $|\psi(0) \rangle$ can be written in terms of energy eigenstates:  $|\psi(0) \rangle= |11...1 \rangle | 22...2 \rangle ...|NN...N \rangle = \sum_{\lambda, a, \alpha} C(\lambda,a, \alpha) |\lambda , a, \alpha \rangle$. (b) Key simplifications arise in the matrix element $\langle \lambda',a',\alpha| Q | \lambda,a,\alpha \rangle$ (which is used to calculate $Q(t)$) since: $\hat{Q} $ is a component of a ``spherical tensor'' for $SU(N)$ (allowing us to make use of the Wigner-Eckart theorem) and has support only on the first $m_1$ sites. (c) The recoupling coefficient is defined by taking the direct product of three irreps $A$, $B$ and $C$, and finding the overlap between two copies of the same irrep found in two ways: by combining $A$ and $B$ first (top), and by combining $B$ and $C$ first (bottom). }
\label{fig:SpinDiffusionDerivation}
\end{figure}

Now we proceed with the calculation for arbitrary $N$, simplifying our notation by dropping hats and vectors. The initial state [see Fig.~\ref{fig:SpinDiffusionDerivation}(a)] can be written as a direct product of spin-symmetric states $| \psi(0) \rangle = \otimes_{j=1}^{m_1}|1 \rangle \otimes_{j=1}^{m_2}|2 \rangle ...\otimes_{j=1}^{m_N}|N \rangle = |\kappa_1,a_1 \rangle |\kappa_2,a_2 \rangle ... |\kappa_N,a_N \rangle$, where $a_i$ labels the particular state in the $\kappa_i \equiv (m_i,0,...0)$ irrep which corresponds to $|i\rangle^{\otimes m_i}$. The product of $\kappa = (m,0,...,0)$ with any irrep $\lambda'$ has no multiplicity \cite{bacon07}: $|\kappa,a\rangle |\lambda',a' \rangle = \sum_{\lambda'', a''} C(\lambda'',a'')| \lambda'', a'' \rangle$, where each irrep $\lambda''$ appears at most once and $C(\lambda'',a'') \equiv \langle \lambda'', a'' | \left( |\kappa,a\rangle |\kappa',a' \rangle \right) $ is a Clebsch-Gordan coefficient. Applying this iteratively, starting from the right, $|\psi(0) \rangle = \sum_{\lambda, a, \alpha} C(\lambda,a, \alpha) |\lambda , a, \alpha \rangle$, where $\alpha$ labels the set of intermediate irreps, $C(\lambda,a,\alpha)$ can be expressed in terms of Clebsch-Gordan coefficients,  and $|\lambda , a, \alpha \rangle$ are orthogonal eigenstates: $H |\lambda , a ,\alpha \rangle = E(\lambda) |\lambda , a, \alpha \rangle$. Note: $a \in 1,2,...,\text{dim}[\lambda_{SU(N)}]$ labels a basis state within the $\lambda$-irrep of $SU(N)$, and each $\alpha$ labels one distinct copy (out of $\text{dim}[\lambda_{S_n}]$ copies) of the $\lambda$-irrep of $SU(N)$ in the Hilbert space $\mathcal{H} = (\mathbb{C}_N)^{\otimes n}$ (all copies of irrep $\lambda$ of $SU(N)$ in $\mathcal{H}$ sit inside a single copy of irrep $\lambda$ of $S_n \times SU(N)$). Therefore: $Q(t) = \sum_{\lambda,\lambda' ,a,a', \alpha} C^*( \lambda', a', \alpha) C( \lambda, a,\alpha) e^{i [E(\lambda') -E(\lambda)]t }  \langle \lambda', a' ,\alpha| Q | \lambda, a, \alpha \rangle$, where we set $\alpha' = \alpha$ since $Q$ has support only on the first $m_1$ spins. We now outline tools to determine the matrix element $\langle \lambda',a',\alpha| Q | \lambda,a,\alpha \rangle$.
 
The states $| \lambda,a,\alpha \rangle$ transform according to matrix irrep $D^{\lambda}$ of $SU(N)$: $ V^{\otimes n} |\lambda,a,\alpha \rangle = \sum_{a'} D^{\lambda}_{a' a}(V)| \lambda,a',\alpha \rangle$. For each $N$, there is a set of single-spin operators which generate $SU(N)$: $\mathbb{\tau}^{\text{adj}} \equiv \{ t_{1}, t_{2} , ..., t_{N^2-1} \}$ which transform according to $D^{\text{adj}}$ (the \textit{adjoint irrep} $\lambda_\text{adj}$): $V^{\otimes n} t_{a} V^{\dagger \otimes n} = \sum_{a'} D^{\text{adj}}_{a' a}(V)t_{a'}$. The set $\{ t_{1}, t_{2} , ..., t_{N^2-1}, \hat{I} \}$ forms a basis for $N\times N$ Hermitian matrices: therefore, any single-atom spin observable can be written as $\hat{q} = c_0 \hat{I} + \sum_{a} c_{a} t_{a}$ for some real constants $c_a$. Therefore $\langle \lambda',a',\alpha| Q | \lambda,a,\alpha \rangle = c_0 + \sum_{a''} c_{a''} \langle \lambda',a',\alpha| T_{a''}^{\text{adj}} | \lambda,a,\alpha \rangle$, where $T_{a}^{\text{adj}} = \sum_{j=1}^{m_1} t_{a\,j}$ and $Q =  \sum_{j=1}^{m_1} |1 \rangle_j \langle 1 |_j \equiv \sum_{j=1}^{m_1} \hat{q}_j$. We now prove a generalization of Eq.~(\ref{UsefulIdentitySU2}) to determine the matrix element $\langle \lambda',a',\alpha'| T_{a''}^{\text{adj}} | \lambda,a,\alpha \rangle$ [see Fig.~\ref{fig:SpinDiffusionDerivation}(b)]. We will need the Wigner-Eckart theorem and recoupling coefficients for $SU(N)$:
\begin{eqnarray}
\label{WignerEckartSUN}
\langle \lambda',a',\alpha' | T^{\lambda''}_{a''}| \lambda,a,\alpha \rangle &=& \sum_{\mathcal{I}} \left( \langle \lambda',a',\mathcal{I} || \lambda'',a'' \rangle | \lambda,a \rangle \right) ~ \langle \lambda',\alpha' || T^{\lambda''} || \lambda,\alpha \rangle_{\mathcal{I}}, \\
\label{recouplingcoefficients}
\left\{ \begin{array}{ccc}
\lambda_A & \lambda_B & \lambda_{AB} \\
\lambda_C & \lambda & \lambda_{BC} \end{array} \right\}_{\mathcal{I}_{AB},\mathcal{I}_C ; \mathcal{I}_{BC},\mathcal{I}_A}  &\equiv& \langle \lambda, a, (\lambda_{AB},\mathcal{I}_{AB},\mathcal{I}_C) |\lambda,a,(\lambda_{BC},\mathcal{I}_{BC},\mathcal{I}_A) \rangle.
\end{eqnarray}
Note that multiplicity $\mathcal{I}$ appears in the Wigner Eckart theorem for $N>2$ [Eq.~(\ref{WignerEckartSUN})], since the tensor product of irreps can include multiple appearances of the same irrep. The recoupling coefficient defined in Eq.~(\ref{recouplingcoefficients}) relates two copies of the same irrep $\lambda$ formed from the tensor product of three irreps: ${\lambda_A}$,  ${\lambda_B}$, and ${\lambda_C}$, but combined in different orders [see Fig.~\ref{fig:SpinDiffusionDerivation}(c)].
To define notation: $\lambda_A$ and $\lambda_B$ are combined to make $\lambda_{AB}$, whose different copies are labeled by $\mathcal{I}_{AB}$, while $\mathcal{I}_{C}$ labels different copies of $\lambda$ when $\lambda_{AB}$ is combined with $\lambda_C$. 

One can decompose $| \lambda, a, \alpha \rangle =\sum_{a_1 ,a_2 } D(a_1 ,a_2 )| \kappa_1,a_1 \rangle | \lambda_2, a_2 \rangle$, where $\lambda_2$ is specified by $\alpha$, and $D \equiv (\langle \kappa_1, a_1 | \langle\lambda_2,  a_2 |) | \lambda, a, \alpha \rangle$. Substituting into $\langle \lambda',a',\alpha| T_{a''}^{\text{adj}} | \lambda,a,\alpha \rangle$ and applying Eq.~(\ref{WignerEckartSUN}) to the first $m_1$ spins:
\begin{eqnarray}
\label{FirstStepUsefulIdentitySUd}
\langle \lambda',a', \alpha| T_{a''}^{\text{adj}} | \lambda,a, \alpha \rangle &=& \langle \kappa_1 || \mathbb{T}^{\text{adj}} || \kappa_1 \rangle ~ \sum_{a_1,a_1',a_2} \left[ (\langle \kappa_1, a_1' | \langle \lambda_2, a_2 |)| \lambda',a' \rangle \right]^* \left[ (\langle \kappa_1, a_1 | \langle \lambda_2, a_2 |)| \lambda,a \rangle \right] \left[ \langle \kappa,a_1' |(| \lambda^{\text{adj}} ,{a''} \rangle | \kappa_1,a_1 \rangle) \right]  \nonumber \\
\label{UsefulIdentitySUd}
&=& \langle \kappa_1 || \mathbb{T}^{\text{adj}} || \kappa_1 \rangle~ \sum_{\mathcal{I}_{1}} \left\{ \begin{array}{ccc}
 \lambda^{\text{adj}} & \kappa_1 & \kappa_1 \\
 \lambda_2 & \lambda' & \lambda \end{array} \right\}_{\mathcal{I}_{1}}^*  \left[ (\langle \lambda^{\text{adj}} ,a''| \langle \lambda,a|)|\lambda',a',\mathcal{I}_1 \rangle \right].
\end{eqnarray}
The second line represents the generalization of Eq.~(\ref{UsefulIdentitySU2}). To derive Eq.~(\ref{UsefulIdentitySUd}), we return to the abstract scenario of three irreps $\lambda_A$, $\lambda_B$ and $\lambda_C$ used to define recoupling coefficients in Eq.~(\ref{recouplingcoefficients}). First write $| \lambda, a, (\lambda_{AB}) \rangle$ as a linear combination of $|\lambda,a,(\lambda_{BC},\mathcal{I}_A) \rangle$ with Eq.~(\ref{recouplingcoefficients}) as coefficients in the special case where $\lambda_B = \lambda_{AB} = \kappa$ (allowing us to drop $\mathcal{I}_{AB}$, $\mathcal{I}_{C}$ and $\mathcal{I}_{BC}$). Rewriting states on both sides as the direct product of states in each of the three subsystems, multiplying by $\left[ \langle \lambda_{BC}',a_{BC} | \left( | \lambda_B,a_B \rangle |  \lambda_C,a_{C} \rangle \right) \right] $, summing over $\lambda_{BC}'$, and using orthogonality gives:
\begin{eqnarray}
\label{UsefulRecouplingRelation}
\sum_{a_{AB},a_B,a_C}\left[ \left(\langle \lambda_{AB},a_{AB}| \langle \lambda_C,a_C|\right)|\lambda,a\rangle \right] \left[ \left(\langle \lambda_A,a_{A}| \langle \lambda_B,a_B|\right)|\lambda_{AB},a_{1AB}\rangle \right] \left[ \langle \lambda_{BC},a_{BC} | \left( | \lambda_B,a_B \rangle |  \lambda_C,a_{C} \rangle \right) \right] &=& \nonumber \\
 \sum_{\mathcal{I}_{A}} \left\{ \begin{array}{ccc}
 \lambda_A & \kappa & \kappa \\
 \lambda_C & \lambda & \lambda_{BC} \end{array} \right\}_{\mathcal{I}_{A}}^*  \left[ \left(\langle \lambda_A,a_A| \langle \lambda_{BC},a_{BC}|\right)|\lambda,a,\mathcal{I}_A \rangle \right]  .
\end{eqnarray}

Using Eq.~(\ref{UsefulIdentitySUd}), the time evolution $T_{a}(t) \equiv \langle \psi(0) | \exp{(i H t)} T_{a} \exp{(-i H t)}|  \psi(0) \rangle$, and therefore $Q(t)$, is written as an efficiently computable sum (containing $poly(n)$ terms \cite{alex11}, each calculated in $poly(n)$ operations):
\begin{eqnarray}
\label{GeneralTimeEvolution}
T_{a}(t)  &=& \langle \kappa_1 || \mathbb{T}^{\text{adj}} || \kappa_1 \rangle~ \sum_{\mathclap{\lambda_1',a_1', \lambda_1,a_1;\alpha}} C^*(\lambda_1', a_1',\alpha ) ~C(\lambda_1, a_1,\alpha ) e^{(i[E(\lambda_1')-E(\lambda_1)]t)}  \\
&& \times   \sum_{\mathcal{I}_{1}} \left\{ \begin{array}{ccc}
 \lambda^{\text{adj}} & \kappa_1 & \kappa_1 \\
 \lambda_2 & \lambda_1' & \lambda_1 \end{array} \right\}_{\mathcal{I}_{1}}  \left[ \langle \lambda_1',a_1',\mathcal{I}_1 |\left( | \lambda^{\text{adj}},j \rangle | \lambda_1,a_1 \rangle \right)  \right] \nonumber .
\end{eqnarray}

The group-theoretic method presented in this Section was crucial for obtaining the analytical result for $SU(2)$ [Eq.~(4) in the main text]. It is also crucial for doing numerics for $SU(N>2)$ for large $n$. However, for sufficiently small n, such as the one shown in Fig.~3, one can do the $SU(N>2)$ numerics using the following simpler method. One first constructs a complete basis of fully symmetric states for the first $m_1$ spins, for the next $m_2$ spins, for the next $m_3$ spins, etc... Then one combines them into a basis for the full system and keeps only those states that have $m_1$ $1$'s, $m_2$ $2$'s, $m_3$ $3$'s, etc... It is straightforward to  evaluate the Hamiltonian in this reduced basis and then numerically exponentiate it to calculate time evolution.

\section{Hamiltonian derivation: atoms with contact interactions} 

In this Section, we derive the Hamiltonian describing identical (bosonic or fermionic) multi-component particles in an infinite square well interacting via $s$-wave interactions. We then specialize to the case of fermionic alkaline-earth atoms and derive Eq.\ (5) in the main text.

Contact interactions between two identical multi-component fermionic (bosonic) atoms are described by the Hamiltonian
\begin{eqnarray}
\hat{H}^{12}_{int} &=& 4 \pi \hbar \omega_\perp \delta(x_1 - x_2) \otimes  \hat{A},
\end{eqnarray}
where the operator $\hat{A}$ only has a physical effect on exchange antisymmetric (symmetric) two-particle internal states because exchange symmetric (antisymmetric) spatial states do not interact. In second quantized form, where $\hat{c}^{\dagger}_{j r}$ creates an atom in internal state $r$ and orbital $\phi_j(x)$ with non-interacting energy $E_j$, and $W_{k' j' j k} = (4 \pi \hbar \omega_\perp) \int_{0}^L dx~ \phi_{k'}(x) \phi_{j'}(x) \phi_{j}(x) \phi_k(x)$. The interaction becomes: $ \hat{H}_{int} = \sum_{j', k', j,  k} W_{ k' j'  j k} \sum_{r', s', r, s}  \langle s', r'| \hat{A} | r,s \rangle ~ \hat{c}^{\dagger}_{j' r'} \hat{c}^{\dagger}_{k' s'} \hat{c}_{j r} \hat{c}_{k s}$.  Specializing to the infinite square well of width $L$, to first order in the interaction, only terms satisfying $(j',k') = (j,k)$ or $(j',k') = (k,j)$ survive. Additionally assuming no multiple occupancies,
we obtain $W_{kjjk} = W_{jkjk} =W \equiv (4 \pi \hbar \omega_\perp)/L$ for $j \neq k$, and the Hamiltonian becomes:
\begin{eqnarray}
\hat{H}&=& \sum_{j , r}  E_j \hat{c}^{\dagger}_{j r} \hat{c}_{j r}  ~+ W \sum_{\mathclap{j ,k }}  \sum_{r', s', r, s} \langle s', r'| \hat{A} | r,s \rangle ~  \left(  \hat{c}^{\dagger}_{j r'} \hat{c}^{\dagger}_{k s'} \hat{c}_{j r} \hat{c}_{k s}  + \hat{c}^{\dagger}_{k r'} \hat{c}^{\dagger}_{j s'} \hat{c}_{j r} \hat{c}_{k s} \right) .  
\end{eqnarray}
Now we specialize to the case focused on in this paper. For fermionic alkaline-earth atoms, $\hat{A}$ cannot depend on nuclear spin; therefore $\hat{A}= \left( a_{ee} |e,e \rangle  \langle e,e | + a_{gg} |g,g \rangle  \langle g,g | + a_{eg}^+ |e,g \rangle_+  \langle e,g |_+ + a_{eg}^- |e,g \rangle_-  \langle e,g |_- \right) \otimes \hat{I}_{Nuclear}$, where $|e,g \rangle_{\pm} =  (|e,g \rangle \pm |g,e \rangle) /\sqrt{2}$ \cite{gorshkov10}. Under these conditions, and applying a strong magnetic field (which to first order in perturbation theory prevents exchanges $|ep,gq \rangle \leftrightarrow |eq,gp \rangle$ for $p \neq q$), 
we obtain Eq.~(5) with
$ U_{1g2g}= U_{2g1g}= U_{gg} \equiv 4 \pi  \omega_\perp a_{gg}/L$, $ U_{1e2e}= U_{2e1e}= U_{ee} \equiv 4 \pi  \omega_\perp a_{ee}/L$, 
$U_{1g1e}=U_{2g2e}= 4 \pi  \omega_\perp a_{eg}^-/M$, 
$U_{1g2e}=U_{2g1e} = 2 \pi  \omega_\perp (a_{eg}^+ + a_{eg}^-)/M$. Recently discovered orbital Feshbach resonances may be used to further tune the values of $U_{1g2e}$ and $U_{2g1e}$ \cite{zhang15d,pagano15,hofer15}.

\section{Experimental Details}

Here we expand upon the experimental considerations section in the main text. The bow tie configuration build-up cavity of attractive magic-wavelength ($\lambda$ =813 nm) beams shown in Fig.~5 in the main text results in orthogonal standing waves in the $x$-$y$ plane, whose intensity maxima are spaced by $\simeq 3$~$\mu$m, with beam waist of 100~$\mu$m at the intersection of the two beams. The build-up cavity will increase the beams' intensity by a factor of $\sim 100$ with a circulating power of 25 W. The resulting 1D trap sites have $\omega_{\perp} \simeq 2 \pi \times  88$~kHz for the initial loading and cooling phase of the experiment.  The (much weaker) longitudinal trapping frequency that results is $\omega_z \simeq 2 \pi \times  880$~Hz. 

As described in the main text, an additional blue-detuned optical potential at 394~nm, the Sr blue magic wavelength, creates sharp caps on the resulting tubes. This potential is formed by a projected image of a Gaussian beam with waist 30~$\mu$m and total power 400~mW screened in the center by a rectangular mask of width $L$ = 10~$\mu$m.

The large $\omega_{\perp}$ enforces a pseudo one-dimensional system as only the lowest radial energy level will be populated. However, the desired condition that $U=(4 \pi a_{gg} \hbar \omega_\perp)/L<3 \hbar^2(\pi /L)^2/M$ is not satisfied with this large $\omega_{\perp}$. After loading into the hybrid red- and blue-detuned optical potential, we propose to ramp the red-detuned optical lattice potentials adiabatically from the 25~W  circulating power to 300~mW, resulting in $\omega_{\perp} \simeq 2 \pi \times 10$~kHz and $\omega_z \simeq 2 \pi \times  100$~Hz. The adiabatic nature of the ramp ensures that the $x$ and $y$ degrees of freedom remain frozen. 

Imperfections on the mask that creates the flat potential and imperfect edges of the trap from the blue-detuned potential contribute to $ \Delta U$. In the following section (Sec.\ \ref{sec:robust}), we give an analytic bound that a harmonic perturbation of frequency $\omega_{z}$ small enough that $M \omega_{z}^2 L^2< \frac{\hbar^2 \pi^2 }{ML^2}$ leads to $ \Delta U/U < 10^{-2} $. Exact diagonalization of the 1D potential confirms that $\Delta U/{U}$ is even less sensitive to $\omega_{z}$: our parameters correspond to $M \omega_{z}^2 L^2 \approx 750  \frac{\hbar^2 \pi^2 }{2ML^2}$, yet $ \Delta U/U$ remains below one percent. The imaging system used to form the potential contributes much more significantly to $\Delta U$. With an imaging point spread function of full width at half maximum (FWHM) of 1~$\mu$m with atoms at 1~$\mu$K, exact diagonalization results in $\Delta U/{U} \lesssim 5$\%. 

Therefore with these parameters, one obtains $U/\hbar =(4 \pi a_{gg} \omega_\perp)/L \approx 2 \pi \times 10$ Hz, and should be able to meet all three of the key requirements stated in the main text with  $\lesssim 20$ atoms in a single tube. In addition, as the pulses in the GHZ protocol should resolve $U$, they should have a sufficiently long duration $\gg 0.1$ s. With additional effort, it should be possible to reach a regime of higher $U$ and $n$ while satisfying these requirements. By shaking the trap during preparation with frequencies low enough to depopulate the lowest $m$ energy orbitals, the restrictions on $L$ and $\omega_\perp$ from the requirement that $ U= (4 \pi a_{gg} \hbar \omega_\perp)/L <3 \hbar^2(\pi /L)^2/M$ is relaxed to $ (4 \pi a_{gg} \hbar \omega_\perp)/L < [(m+2)^2 - (m+1)^2] \hbar^2(\pi /L)^2/M$. Decreasing the ratio between the spatial imperfections of the potential and $L$ will reduce $\Delta U/{U}$. For example, reducing the FWHM of the point spread function in our numerical calculations described above from 1~$\mu$m to 0.5~$\mu$m yields $\Delta U/{U} < 2$\%. Approaches for creating subwavelength potentials can also be envisioned \cite{jendrzejewski15}.

Beyond the three key requirements given in the main text, there are a number of other considerations which we now address. Taking a typical recombination rate constant $K_3 \approx 10^{-28}$ cm$^6/$s for $n = 20$ particles, it should take approximately 1 second before a single particle is lost.  This loss time is 10 times longer than the coherent interaction time $2 \pi \hbar/U$, a ratio that 
is comparable (or even superior) to the ratio of the decoherence time to the spin-spin interaction time in superexchange-based systems \cite{trotzky08,brown15}. Tunneling between the tubes is negligible due to the large $3$ $\mu$m spacing between tubes. 
The approximate magnitude of $p$-wave terms involving occupied orbitals $j$ and $k$ is $\pi^2 (j^2+k^2)(b_{gg}/L)^2 (b_{gg}/a_{gg}) \,U$, where $b_{gg}^3$ is the scattering volume for $p$-wave interactions. This remains small for $j,k <300$, taking $ b_{gg} \approx 3.9$ nm \cite{zhang14} for $^{87}$Sr. Vector and tensor light shifts \cite{boyd07c} in principle break $SU(N)$ symmetry, but tensor polarizability in our system is negligible, while vector shifts can be avoided with the use of linear polarization \footnote{Small deviations from linear polarization will play a more significant role for the ${}^3$P$_0$ state than for the ${}^1$S$_0$ state because of the larger vector polarizability of the former. However,  the ${}^3$P$_0$ state is only used in the GHZ protocol where a vector shift is indistinguishable from a slight change in  the value of the applied magnetic field.}.
Specifically, to ensure any breaking of the $SU(N)$ symmetry is far below a level which could affect our proposal, beam circularity of below a few percent should be sufficient. An appropriate choice of linear polarization of the blue-detuned beam will ensure minimal longitudinal field components (and hence minimal circularity) induced by imaging the mask.
  
\section{Robustness to imperfections \label{sec:robust}}

In this Section, we consider deviation from a perfect infinite square-well potential $V(x)$. For simplicity, we consider the case in which all atoms are in the ground electronic state. The interaction Hamiltonian Eq.~(1) in the main text becomes: $\hat{H}' = -\sum_{j < k} U_{jk} \hat{s}_{jk}$, where $U_{jk} = (UL/2) \int \phi_j^2(x) \phi_k^2(x) dx$, and $\phi_j(x)$ is a single-particle orbital, which is a sine function in the ideal case. As $\hat{H}' $ is a weighted sum of terms $\hat{s}_{jk}$ and therefore has $SU(N)$ symmetry, it cannot mix states in different $\vec{\lambda}$-subspaces. However as $\hat{H}' $ does not exhibit $S_n$ symmetry, the $\vec{\lambda}$ subspace does not have a single energy - but breaks into $D(\vec{\lambda})$ energy subspaces, $D(\vec{\lambda})$ is the dimension of the $\vec{\lambda}$ irrep of $S_n$. We write the eigenenergies of $\hat{H}' $ as $E'(\vec{\lambda},b)$, with $b$ labeling distinct energies.

Provided that the inhomogeneity in $U_{jk}$ is small, i.e.\ that $|U_{jk}-U| \ll U$, the energy splittings $E'(\vec{\lambda},b)$ within each $\vec{\lambda}$ subspace will be small compared to energy separations between different $\vec{\lambda}$ subspaces. Exact determination of $E(\vec{\lambda},b)$ can be carried out by projecting $\hat{H}'$ onto the $\vec{\lambda}$ subspace and solving the resulting matrix equation,  which is computationally difficult as the matrices have dimension $O(\exp(n))$. Here we are satisfied with an indication of the magnitude of deviation from the ideal energy eigenvalues. We seek the offset: $\Delta E(\vec{\lambda}) \equiv \frac{1}{D(\vec{\lambda})} \sum_{b=1}^{D(\vec{\lambda})} \left[E'(\vec{\lambda},b)-E(\vec{\lambda}) \right] $ and the variance: $\sigma^2(\vec{\lambda}) \equiv \frac{1}{D(\vec{\lambda})} \sum_{b=1}^{D(\vec{\lambda})} \left[ \Delta E(\vec{\lambda},b) - \Delta E(\vec{\lambda}) \right]^2$. Defining $E(\vec{\lambda}_0) = -U n(n-1)/2$,  where $\vec{\lambda}_0=(n,0,0,..,0)$, one can show that
\begin{equation} 
\Delta E(\vec{\lambda}) = - \left( \frac{E(\vec{\lambda})}{E(\vec{\lambda}_0)} \right) \sum_{j < k}  (U_{jk}-U).
\end{equation}
Note that $\left| \frac{E(\vec{\lambda})}{E(\vec{\lambda}_0)} \right| \leq 1$ for all $\vec{\lambda}$. The main technical lemma used to prove this is that for any operator $\hat{O}$, 
\begin{equation} \sum_{b=1}^{D(\vec{\lambda})} \langle \vec{\lambda}, b| \hat{O} | \vec{\lambda},b \rangle = \frac{D(\vec{\lambda})}{n!} \sum_{\sigma \in S_n}  \langle \vec{\lambda}, b'| \sigma^{-1}\hat{O} \sigma | \vec{\lambda},b' \rangle,
\end{equation}
where the latter sum is over all permutations $\sigma$ in the symmetric group $S_n$.
Modeling $U_{jk}$ as a set of $n(n-1)/2$ independent random variables with mean $U$, one can similarly show that
\begin{equation} \sigma^2(\vec{\lambda}) = \left[1- \left( \frac{E(\vec{\lambda})}{E(\vec{\lambda}_0)} \right)^2 \right] \sum_{j < k} \langle (U_{jk}-U)^2 \rangle,
\end{equation}
where $\langle \rangle$ indicates that we have taken the ensemble average over realizations \footnote{It is not necessary to do this -- one can calculate the exact expression without taking an ensemble average, but it is quite complicated, and all we seek is an approximate indication of how much spreading to expect for each subspace.} of $\Delta U_{jk}$, which simply allows us to set $\langle \Delta U_{jk} \Delta U_{j'k'} \rangle = 0$ where $j,k \neq j',k'$. These results indicate that the deviations in energy levels from those for the exact case caused by inhomogeneity in $U_{jk}$ generically behave as $\sim n \Delta U$. This is because, to estimate $\Delta E(\vec{\lambda})$, we assume that $\sum_{j < k}  (U_{jk}-U)$ is the sum of $n(n-1)/2$ uncorrelated positive and negative terms each of magnitude $ \sim \Delta U$, and similarly for the variance $\sigma^2(\vec{\lambda})$, except all terms are positive. We therefore expect that, in order to see $p$ revivals of the kind shown in Fig.\ 3 of the main text, we need to pick up small phase errors $n \Delta U t \lesssim 1$ over time $t \sim p/U$, which corresponds to $\Delta U/U \lesssim 1/(n p)$.

However, note that most symmetric $\vec{\lambda}$ subspaces (which have $E(\vec{\lambda})/E(\vec{\lambda}_0)$ close to unity), experience less splitting due to inhomogeneity in $U_{jk}$, although they do experience an overall shift. For the GHZ protocol described in the main text, the $\vec{\lambda}$ subspaces involved are  $(n,0)$, $(n-1,1)$ and $(n-2,2)$, which will shift relative to one another under inhomogeneity in $U_{jk}$ by an amount independent of $n$ for large $n$.

To obtain some concrete estimates of the effects of an imperfect square-well potential, we consider the following example: a perfect square well, plus an additional harmonic perturbing potential $V_1(x) = \alpha x^2$ (which in effect ``rounds off" the boundary of the well somewhat). With first-order corrections, the single-particle wave functions $\phi_j(x)$ are
\begin{eqnarray} 
\phi_j(x)  &=&  \sqrt{\frac{2}{ L}}\sin{(j \pi x /L)} + \frac{8}{\pi^2} \left( \alpha L^2 / \frac{\hbar^2 \pi^2}{2 M L^2}  \right) \sum_{k ~ k \neq j } \frac{j k (-1)^{j+k}}{(j^2-k^2)^3} ~\sqrt{\frac{2}{ L}}  \sin{(k \pi x /L)}.
\end{eqnarray}

Substitution into $U_{jk} = UL\int \phi_j^{2}(x) \phi_k^{2}(x) dx$ yields exact expressions for the first order corrections to $U$, which (for all $j$ and $k$) satisfy: $ |U_{jk} - U| < 10^{-2} \left( \alpha L^2 / \frac{\hbar^2 \pi^2 }{2ML^2} \right)  U + O(\alpha^2) $. The inhomogeneity is therefore strictly less than one percent if the magnitude of the perturbation is approximately of the same order as the characteristic energy of the square well. The size of the deviations fall off at the fourth power of $j,k$, such that for ensembles of atoms, $\Delta U$ is typically much better than this bound suggests.

\section{GHZ state preparation} 

In this Section, we present the details behind the GHZ state preparation protocol and explain how $m$ GHZ states can be prepared when $N \geq 2^m$.

The state $|A\rangle =|1g\,1g...1g \rangle$ has energy $E_A =n B_g$. The state $|B\rangle = |\{1e\,1g...1g \} \rangle$ lies in the same energy manifold as the state $(|1g \,1e \rangle -|1e \,1g \rangle) |1g...1g \rangle$, which has energy $E_B = \omega_{eg} + (n-1)B_g + B_e +[(n-1) -(-1)]U_{1g\,1e}
$. Similarly, $|C\rangle = |\{1e\,1e...1g \} \rangle$ has the same energy as $(|1g \,1e \rangle -|1e \,1g \rangle)(|1g \,1e \rangle -|1e \,1g \rangle) |1g...1g \rangle$, with energy $E_C = 2\omega_{eg} + (n-2)B_g + 2B_e +[2(n-2) -(-2)]U_{1g\,1e}
$. Driving with frequency $(E_B- E_A)$ forms an effective two-level system: $\{|A\rangle \leftrightarrow |B\rangle \not\leftrightarrow |C \rangle \}$ since $(E_B- E_A) = \omega_{eg} -B_g + B_e +nU_{1g\,1e} \neq (E_C-E_B) =  \omega_{eg} -B_g + B_e +(n-2)U_{1g\,1e} $. 
Now we explain why transition $|A\rangle \rightarrow |D\rangle  \equiv |2g\,2g...2g\rangle $ occurs, while the transition 
$ |B\rangle \not\rightarrow |x\rangle$ is blocked for any energy eigenstate $|x\rangle$. First note that the transition $|A\rangle \rightarrow |D\rangle$ actually passes through a ladder of intermediate energy eigenstates: $|A\rangle \equiv |1g\,1g...1g \rangle \rightarrow |\mathcal{S}\{2g\,1g...1g\} \rangle  \rightarrow |\mathcal{S}\{2g\,2g...1g\} \rangle \rightarrow ...\rightarrow |2g\,2g...2g\rangle \equiv |D\rangle$, where $\mathcal{S}$ symmetrizes its argument. Each state in the ladder has energy $2B_g$ more than the last, and is connected to the previous through the operator $\hat{P} = \sum_{j} ( |2g \rangle_j \langle 1g|_j +| 2e \rangle_j \langle 1e|_j  )$, which is applied as a pulse with frequency $2B_g$. To show that $ |B\rangle$ does not transition to any other state under the action of this pulse, we must prove that \textit{there exists no state $|x\rangle$ such that $\hat{H}|x\rangle = (E_B+ 2B_g)|x\rangle$ and $\langle x | \hat{P}|B\rangle \neq 0$.} We will assume that $n>2$, $B_e \neq B_g$ and either $|U_{gg}|>0$ or $|U_{eg}|>0$.

Our proof has the following structure: we find four orthonormal states such that $ \hat{P}|B\rangle \in \text{span} \{|\phi_1\rangle, |\phi_2\rangle, |\phi_3\rangle, |\phi_4\rangle \} \equiv \mathcal{H}_0$, where subspace $\mathcal{H}_0$ is closed under the action of $\hat{H}$ (i.e. for all $|\psi\rangle \in \mathcal{H}_0, ~\hat{H} |\psi \rangle \in \mathcal{H}_0$). Any eigenstate $|x\rangle$ of $\hat{H}$ coupled to $|B\rangle$ through $ \hat{P}$ must be in $\mathcal{H}_0$, but we show the four eigenvalues $E_i$ of $\hat{H}$ in $\mathcal{H}_0$ satisfy $E_i \neq (E_B- 2B_g)$.  

To complete the proof, we must present $\{ |\phi_1\rangle, |\phi_2\rangle, |\phi_3\rangle,|\phi_4\rangle \}$ explicitly, and show that $E_i \neq (E_B- 2B^g)$ for all four eigenstates ($i=1,2,3,4$). Without loss of generality, take $|B\rangle = (|1g \,1e \rangle -|1e \,1g \rangle) |1g...1g \rangle$, thus $\hat{P}|B\rangle = \sqrt{2(n-2)}|\phi_1\rangle +\sqrt{2}|\phi_3\rangle  + \sqrt{2}|\phi_4\rangle$, where $|\phi_1\rangle \equiv \frac{1}{\sqrt{2 (n-2)}} (|1g \,1e \rangle -|1e \,1g \rangle) |\mathcal{S} \{1g2g...1g \}\rangle$, $|\phi_2\rangle \equiv \frac{1}{\sqrt{2}} (|2g \,1e \rangle -|1e \,2g \rangle) |1g1g...1g \rangle$, $|\phi_3\rangle \equiv \frac{1}{\sqrt{2 (n-2)}} (|1g \,2g \rangle -|2g \,1g \rangle) |\mathcal{S} \{1g1e...1g \}\rangle$, and $|\phi_4\rangle \equiv \frac{1}{\sqrt{2}} (|1g \,2e \rangle -|2e \,1g \rangle) |1g1g...1g \rangle$ (note that $|\phi_4\rangle$ is an energy eigenstate). $\hat{H}$ is closed on subspace $\mathcal{H}_0$ and takes the form:
\begin{equation}
\label{SubspaceHamiltonian}
\hat{H} =(E_B -2B_g)  +\left( \begin{array}{cccc}
 0 & - \sqrt{n-2}U_{gg}  & -U_{ge} & 0 \\
- \sqrt{n-2}U_{gg}  & (n-2)U_{gg}  & \sqrt{n-2}U_{ge} & 0 \\
 -U_{ge} & \sqrt{n-2}U_{ge}  & (n -1)U_{gg} -U_{ge} & 0 \\
 0 & 0  & 0 & 2(B_g-B_e)  \end{array}  \right).
\end{equation}
The matrix written explicitly in Eq.~(\ref{SubspaceHamiltonian}) can be shown 
to have non-zero determinant (and therefore no vanishing eigenvalues)
provided $n>2$, $B_e \neq B_g$ and either $|U_{gg}|>0$ or $|U_{eg}|>0$, which completes our proof.

In the main text, we note that for integer $m$ such that $N \geq 2^m$, it is possible to create $m$ GHZ states. We describe the procedure here in more detail for $m=2$. First create a regular GHZ state as described in the main text $(|1g1g..1g\rangle + |2g2g..2g\rangle)$ from initial state $|1g...1g\rangle$. Then, apply pulse 1 of two different frequencies to $|1g1g..1g\rangle$ and to $|2g2g..2g\rangle$, resulting in $(|1e1g..1g\rangle +|1g1g..1g\rangle + |2e2g..2g\rangle  + |2g2g..2g\rangle)$.  Now, instead of applying pulse 2, apply a pulse which implements $|p \rangle \mapsto |p+2\rangle$ (for $p=1,2$), but only to atoms in a many-body state containing no $e$ atoms. The resulting state is $(|1e1g..1g\rangle +|3g3g..3g\rangle + |2e2g..2g\rangle  + |4g4g..4g\rangle)$. Finally, apply pulse 3 of two different frequencies to yield $(|1g1g..1g\rangle + |2g2g..2g\rangle  +|3g3g..3g\rangle + |4g4g..4g\rangle)$. This is precisely equivalent to two GHZ states, which can be seen by defining the basis $ \{ \ket{\Downarrow \Downarrow} , \ket{\Downarrow \Uparrow}  ,\ket{\Uparrow \Downarrow}  ,\ket{\Uparrow \Uparrow}  \equiv \{ |1 \rangle, | 2 \rangle , | 3 \rangle, |4 \rangle \}  \}$. Then $(|11..1\rangle + |22..2\rangle + |33..3\rangle + |44..4\rangle) = (\ket{\Downarrow \Downarrow .. \Downarrow } + \ket{\Uparrow \Uparrow .. \Uparrow }) (\ket{\Downarrow \Downarrow .. \Downarrow} + \ket{\Uparrow \Uparrow .. \Uparrow})$. The process could be continued, where in the $i$th iteration, the second pulse involves $|p \rangle \mapsto |p+2^i\rangle$ (for $p = 1,2,3...2^i$).  

\end{widetext}

\bibliography{SUNrefs}

\end{document}